%% file: cosmo.tex
\documentclass[letterpaper]{aa520}
\usepackage{amsmath}
\usepackage{natbib}
\usepackage{graphicx}
\usepackage{txfonts}
\usepackage{color}
\usepackage{deluxetable}

\newcommand{\om}{\Omega_{\rm M}}
\newcommand{\ol}{\Omega_\Lambda}
\newcommand{\ot}{\Omega_{\rm tot}}
\newcommand{\ok}{\Omega_{\rm k}}
\newcommand{\gme}{g_M}
\newcommand{\rme}{r_M}
\newcommand{\ime}{i_M}
\newcommand{\zme}{z_M}

\begin{document}
             
\title{The Supernova Legacy Survey: Measurement of $\om$, $\ol$ and $w$ from
the First Year Data Set
\thanks{
Based on observations obtained with MegaPrime/MegaCam, a joint project
of CFHT and CEA/DAPNIA, at the Canada-France-Hawaii Telescope (CFHT)
which is operated by the National Research Council (NRC) of Canada,
the Institut National des Sciences de l'Univers of the Centre National
de la Recherche Scientifique (CNRS) of France, and the University of
Hawaii. This work is based in part on data products produced at the
Canadian Astronomy Data Centre as part of the Canada-France-Hawaii
Telescope Legacy Survey, a collaborative project of NRC and CNRS.
Based on observations 
obtained at the European Southern Observatory using the Very Large Telescope
on the Cerro Paranal (ESO Large Programme 171.A-0486).
Based on observations (programs GN-2004A-Q-19, GS-2004A-Q-11, GN-2003B-Q-9, and
GS-2003B-Q-8) obtained at the Gemini Observatory, which is
operated by the Association of Universities for Research in Astronomy,
Inc., under a cooperative agreement with the NSF on behalf of the
Gemini partnership: the National Science Foundation (United States),
the Particle Physics and Astronomy Research Council (United Kingdom),
the National Research Council (Canada), CONICYT (Chile), the
Australian Research Council (Australia), CNPq (Brazil) and CONICET
(Argentina).
Based on observations obtained at the W.M. Keck
Observatory, which is operated as a scientific partnership among the
California Institute of Technology, the University of California and the
National Aeronautics and Space Administration. The Observatory was made
possible by the generous financial support of the W.M. Keck Foundation.
}
}
\titlerunning{SNLS 1st Year Data Set}

\author{
P.~Astier\inst{1},
J.~Guy\inst{1},
N.~Regnault\inst{1},
R.~Pain\inst{1},
E.~Aubourg\inst{2,3},
D.~Balam\inst{4},
S.~Basa\inst{5},
R.G.~Carlberg\inst{6},
S.~Fabbro\inst{7},
D.~Fouchez\inst{8},
I.M.~Hook\inst{9},
D.A.~Howell\inst{6},
H.~Lafoux\inst{3},
J.D.~Neill\inst{4},
N.~Palanque-Delabrouille\inst{3},
K.~Perrett\inst{6},
C.J.~Pritchet\inst{4},
J.~Rich\inst{3},
M.~Sullivan\inst{6},
R.~Taillet\inst{1,10},
G.~Aldering\inst{11},
P.~Antilogus\inst{1},
V.~Arsenijevic\inst{7},
C.~Balland\inst{1,2},
S.~Baumont\inst{1,12},
J.~Bronder\inst{9},
H.~Courtois\inst{13},
R.S.~Ellis\inst{14},
M.~Filiol\inst{5},
A.C.~Gon\c{c}alves\inst{15},
A.~Goobar\inst{16},
D.~Guide\inst{1},
D.~Hardin\inst{1},
V.~Lusset\inst{3},
C.~Lidman\inst{12},
R.~McMahon\inst{17},
M.~Mouchet\inst{15,2},
A.~Mourao\inst{7},
S.~Perlmutter\inst{11,18},
P.~Ripoche\inst{8},
C.~Tao\inst{8},
N.~Walton\inst{17}
}
\institute{
LPNHE, CNRS-IN2P3 and Universit\'{e}s Paris VI \& VII, 4 place Jussieu,
75252 Paris Cedex 05, France
\and APC, Coll\`ege de France, 11 place Marcellin Berthelot, 75005 Paris, 
France
\and DSM/DAPNIA, CEA/Saclay, 91191 Gif-sur-Yvette Cedex, France
\and Department of Physics and Astronomy, University of Victoria, 
PO Box 3055, Victoria, BC VSW 3P6, Canada
\and LAM, CNRS, BP8, Traverse du Siphon, 13376 Marseille Cedex 12, France
\and Department of Astronomy and Astrophysics, University of Toronto, 
60 St. George Street, Toronto, ON M5S 3H8, Canada
\and CENTRA-Centro M. de Astrofisica and Department of Physics, IST, Lisbon, 
Portugal
\and CPPM, CNRS-IN2P3 and Universit\'e Aix-Marseille II, Case 907, 
13288 Marseille Cedex 9, France
\and University of Oxford Astrophysics, Denys Wilkinson Building, Keble Road,
Oxford OX1 3RH, UK
\and Universit\'e de Savoie, 73000 Chambery, France
\and LBNL, 1 Cyclotron Rd, Berkeley, CA 94720, USA
\and ESO, Alonso de Cordova 3107, Vitacura, Casilla 19001, Santiago 19, Chile
\and CRAL, 9 avenue Charles Andre, 69561 Saint Genis Laval cedex, France 
\and California Institute of Technology, Pasadena, California, USA
\and LUTH,UMR 8102, CNRS and Observatoire de Paris, F-92195 Meudon, France
\and Department of Physics, Stockholm University, Sweden
\and IoA, University of Cambridge, Madingley Road,
Cambridge, CB3 0EZ, UK
\and Department of Physics, University of California Berkeley, Berkeley, CA 94720, USA
} 

\authorrunning{P. Astier et al, SNLS Collaboration}                         
\offprints{astier\@@in2p3.fr}

\date{Received Month DD, YYYY; accepted Month DD, YYYY}

\abstract{We present distance measurements to 71 high redshift type
  Ia supernovae discovered during the first year of the 5-year
  Supernova Legacy Survey (SNLS). These events were detected and their
  multi-color light-curves measured using the MegaPrime/MegaCam instrument at
  the Canada-France-Hawaii Telescope (CFHT), by repeatedly imaging four one-square degree fields in 
  four bands. Follow-up spectroscopy was performed at
  the VLT, Gemini and Keck telescopes to confirm the nature of the
  supernovae and to measure their redshift.  With this data set, we 
  have built a Hubble diagram extending to $z=1$, with all distance
  measurements involving at least two bands. Systematic uncertainties
  are evaluated making use of the multi-band photometry obtained
  at CFHT. Cosmological fits to this first year SNLS
  Hubble diagram give the following results : $\om = 0.263 \pm
  0.042\;(stat) \pm 0.032\;(sys)$ for a flat $\Lambda$CDM model; and $w =
  -1.023 \pm 0.090\;(stat) \pm 0.054\;(sys)$ for a flat cosmology with
  constant equation of state $w$ when combined with the  
  constraint from the recent Sloan Digital Sky Survey  measurement of 
baryon acoustic oscillations.

}
\maketitle

\section{Introduction}

The discovery of the acceleration of the Universe stands as a 
major breakthrough 
of observational cosmology.
Surveys of cosmologically distant Type Ia supernovae
\citep[SNe~Ia;][]{Riess98b,Perlmutter99} indicated the presence of a
new, unaccounted-for ``dark energy'' that opposes the self-attraction
of matter and causes the expansion of the Universe to accelerate. When
combined with indirect measurements using cosmic microwave background
(CMB) anisotropies, cosmic shear and studies of galaxy clusters, a
cosmological world model has emerged that describes the Universe as
flat, with about 70\% of its energy contained in the form of this cosmic
dark energy (see for example~\citealt{Seljak05}).

Current projects aim at directly probing the nature of the dark energy 
via a determination of
its equation of state parameter -- the pressure to energy-density ratio -- $w
\equiv p_X/\rho_X$, which  
also defines the time dependence of the dark energy density:
$\rho_X \sim a^{-3(1+w)}$, where $a$ is the scale factor.
Recent constraints on $w$ \citep{Knop03,Tonry03,Barris04,Riess04} are
consistent with a very wide range of Dark Energy models.  Among them,
the historical cosmological constant ($w=-1$) is $10^{120}$ to
$10^{60}$ smaller than plausible vacuum energies predicted by fundamental
particle theories. It also cannot explain why matter and dark energy
have comparable densities today. ``Dynamical $\Lambda$'' models have
been proposed (quintessence, k-essence) based on speculative field
models, and some predict values of $w$ above -0.8 -- significantly
different from -1.  Measuring the average value of $w$ with a
precision better than 0.1 will permit a discrimination between the
null hypothesis (pure cosmological constant, $w=-1$)
and some dynamical dark energy models.

Improving significantly over current SN constraints on the dark
energy requires a ten-fold larger sample (i.e. o(1000) at $0.2<z<1.$,
where $w$ is best measured), in order to
significantly improve on statistical errors but also, most importantly,
on systematic uncertainties. 
The traditional method of measuring
distances to SNe~Ia involves different types of observations at
about 10 different epochs spread over nearly 3 months: discovery via
image subtraction, spectroscopic identification, and photometric
follow-up, usually on several telescopes.  
Many objects are lost or poorly measured in this process due to the effects of
inclement weather during the follow-up observations, and the analysis often
subject to largely unknown systematic uncertainties due to the use of various 
instruments and telescopes. 

The Supernova Legacy Survey (SNLS)\footnote{see
\texttt{http://cfht.hawaii.edu/SNLS/}} was designed to improve
significantly over the traditional strategy as follows: 1) discovery
and photometric follow-up are performed with a wide field imager used
in ``rolling search'' mode, where a given field is observed every
third to fourth night as long as it remains visible; 2) service
observing is exploited for both spectroscopy and imaging, reducing the
impact of bad weather. Using a single imaging instrument to observe
the same fields reduces photometric systematic
uncertainties; service observing optimizes both the yield of spectroscopic
observing time, and the light-curve sampling.

In this paper we report the progress made, and the cosmological results
obtained, from analyzing the first year of the SNLS. We present the data 
collected, the precision achieved both from improved statistics and better
control of systematics,    
and the potential of the project to further reduce and control 
systematic uncertainties on cosmological parameters.
Section~\ref{section:survey} describes the imaging and spectroscopic surveys
and their current status.  Sections~\ref{section:data_reduction} and
\ref{section:photometric_calibration} present the data reduction and
photometric calibration. The light-curve fitting method, the SNe samples
and the cosmological analysis are discussed in 
Section~\ref{section:snIa_lc_cosmo}.  A comparison of the nearby and
distant samples used in the cosmological analysis is performed in
Section~\ref{section:comparison} and the systematic uncertainties are
discussed in Section~\ref{section:systematic_uncertainties}.

\section{The Supernova Legacy Survey}
\label{section:survey}

The Supernova Legacy Survey is comprised of two components: an imaging
survey to detect SNe and monitor their light-curves, and a 
spectroscopic program to confirm the nature of the candidates and
measure their redshift.

\subsection{The imaging survey}
\label{subsec:imaging-survey}

The imaging is taken as part of the deep component of the CFHT Legacy
Survey~\citep{cfhtls} using the one square degree imager,
MegaCam~\citep{MegacamPaper}.  In total, CFHTLS has been allocated 
474 nights
over 5 years and consists of 3 surveys: a very wide shallow survey
(1300 square degrees), a wide survey (120 square degrees) and a deep
survey (4 square degrees). The 4 pointings of the deep survey
are evenly distributed in right ascension
(Table~\ref{table:deep_fields}).  The observations for the deep survey
are sequenced in a way suitable for detecting supernovae and measuring
their light-curves: in every lunation in which a field is visible,
it is imaged at five equally spaced epochs during a MegaCam run
(which lasts about 18 nights). Observations are taken in a combination of
$\rme$, $\ime$ plus $\gme$ or $\zme$ filters (the MegaCam filter set;
see Section~\ref{section:photometric_calibration}) depending on the phase of
the moon. Each field is observed for 5 to 7 consecutive lunations.
Epochs lost to weather on any one night remain in the queue until the
next clear observing opportunity, or until a new observation in the
same filter is scheduled.

During the first year of the survey, the observing efficiency was
lower than expected and the nominal observation plan could not always
be fulfilled.  The scheduled $\ime$ exposures (3 $\times$ 3600~s plus 
2 $\times$ 1800s per lunation)
 and $\rme$ exposures (5 epochs x 1500 s) were usually
acquired.  Assigned a lower priority, $\gme$ and $\zme$ received less
time than originally planned: on average only 2.2 epochs of 1050~s
were collected per lunation in $\gme$, and 2 epochs of 2700 s in
$\zme$; for the latter, the average ignores the D2 field and the D3
field in 2003, for which only fragmentary observations were
obtained in $\zme$.  With efficiency
ramping up, $\gme$ and $\zme$ approached their nominal rate in May
2004, and since then
the nominal observation plan (detailed in \citealt{Sullivan05}) is 
usually completed. 

\begin{table}
\centering
\begin{tabular}{ccc|c}
 Field &  RA(2000)    & Dec (2000)  &  E(B-V) (MW)    \\
\hline
D1     &  02:26:00.00 & -04:30:00.0 & 0.027\\
D2     &  10:00:28.60 & +02:12:21.0 & 0.018\\
D3     &  14:19:28.01 & +52:40:41.0 & 0.010\\
D4     &  22:15:31.67 & -17:44:05.0 & 0.027\\
\hline
\end{tabular}
\caption{Coordinates and average Milky Way extinction (from \citealt{Schlegel98}) of 
fields observed by the Deep/SN component of the CFHTLS.\label{table:deep_fields}       
}
\end{table}

Observations and real-time pre-processing are performed by the CFHT
staff using the Elixir reduction pipeline \citep{Magnier04}, with the
data products immediately available to the SN search teams.  We have
set up two independent real-time pipelines which analyze
these pre-processed images.  The detection of new candidates is
performed by subtracting a ``past" image to the current images,
where the past-image is constructed by stacking previous
observations of the same field.  The key element of these pipelines is
matching the point spread function of a new exposure to the 
past-image.  This is done using the Alard algorithm \citep{Alard98,Alard00}
for one of the pipelines, and using a non-parametric approach for the
other.  
New candidates are detected and measured on the subtraction images;
detections are matched to other
detections in the field, if any. 
One of the pipelines processes all bands on an equal footing, the
other detects in the $\ime$ band (which is deep enough for trigger
purposes) and measures fluxes in the other bands. 
The two candidate lists are merged after each epoch and typically have
an overlap greater than 90\% for $\ime(AB)<24.0$ after two epochs in a
dark run. 
The reasons for one candidate being found by only one
pipeline are usually traced to different masking strategies or
different handling of the CCD overlap regions. 

\subsection{Spectroscopic follow-up
\label{section:spectro_follow_up}}

Spectroscopy is vital in order to obtain SN redshifts, and to
determine the nature of each SN candidate.  This requires observations
on 8-10 meter class telescopes due to the faintness of these distant
supernovae.  Spectroscopic follow-up time for the candidates presented
in this paper was obtained at a variety of telescopes during the
Spring and Fall semesters of 2003 and the Spring semester of 2004.
The principle spectroscopic allocations were at the European Southern
Observatory Very Large Telescope (program ID $<$171.A-0486$>$; 60
hours per semester), and at Gemini-North and South (Program-IDs:
GN-2004A-Q-19, GS-2004A-Q-11, GN-2003B-Q-9, and GS-2003B-Q-8; 60 hours
per semester).  
Spectroscopic time was also obtained at Keck-I and Keck-II (3
nights during each Spring semester) as the D3 field
cannot be seen by VLT or Gemini-South.  Further complementary
spectroscopic follow-up observations were also obtained at Keck-I (4
nights in each of 2003A, 2003B and 2004A) as part of a detailed study
of the intermediate redshift SNe in our sample 
(\citeauthor{Ellis05}, in prep.).

Most of the observations are performed in long-slit mode. 
The detailed
spectroscopic classification of these candidates is discussed
elsewhere
(see \citealt{Howell05} and \citeauthor{Basa05}, in prep.). In
summary, we consider two classes of events (see \citealt{Howell05} for
the exact definitions): secure SNe~Ia events ("SN~Ia"), 
and probable Ia events (``SN~Ia*''), for
which the spectrum matches a SN~Ia better than any other type, but
does not completely rule out other possible interpretations.  All
other events which were not spectroscopically identified as SN~Ia or
SN~Ia* were ignored in this analysis.

The
imaging survey still delivers more variable candidates than can
actually be observed spectroscopically. Hence, an accurate ranking of 
these candidates
for further observations is essential. This ranking is performed to
optimize the SN~Ia yield of our allocations. Our method uses both a
photometric selection tool \citep[discussed in][]{Sullivan05} which
performs real-time light-curve fits to reduce the contamination of
core-collapse SNe, and a database of every variable object ever
detected by our pipelines to remove AGN and variable stars which 
are seen to vary repeatedly in
long-timescale data sets (more than one year).

SN~Ia candidates fainter than $\ime = 24.5$ (likely at $z>1$) and
those with very low percentage increases over their host galaxies
(where identification is extremely difficult -- see
\citealt{Howell05}) are usually not observed. With the real-time 
light-curve fit technique, approximately 70\% of our candidates turned out to
be SNe~Ia.  The possible biases associated with this selection were
studied in \cite{Sullivan05} and found to be negligible.

\subsection{The first year data set}
\label{section:first_year}

The imaging survey started in August 2003 after a few months of
MegaCam commissioning. (Some SN candidates presented here were detected
during the commissioning period.)
This paper considers candidates with maximum light
up to July ${\rm 15^{th}}$ 2004, corresponding approximatively to a full 
year of
operation. During this time frame, which includes the ramping-up period of the
CFHTLS, about 400 transients were detected, 
142 spectra were acquired: 20
events were identified as Type II supernovae, 9 as AGN/QSO, 4 as SN~Ib/c, and
91 events were classified as SN~Ia or SN~Ia*.  The 18 remaining events have
inconclusive spectra.  Table \ref{table:wholesnlsIa} gives the 91
objects identified as SN~Ia or SN~Ia* during our first year of
operation.

\section{Data reduction}
\label{section:data_reduction}

\subsection{Image preprocessing}

At the end of each MegaCam run, the images are pre-processed again at CFHT
using the Elixir pipeline \citep{Magnier04}. This differs from the
real-time reduction process described 
in Section~\ref{subsec:imaging-survey},
in that master flat-field images
and fringe-correction frames are constructed from all available data
from the entire MegaCam run (including PI data). The Elixir process
consists of flat-fielding and fringe subtraction, with an approximate
astrometric solution also derived. 
Elixir provides reduced data which has a uniform photometric
response across the mosaic (at the expense of a non-uniform sky background).  
This ``photometric flat-field'' correction is constructed
using exposures with large dithers obtained on dense stellar fields.

The SNLS pipelines then associate a weight map with each Elixir-processed
image (i.e. each
CCD from a given exposure) from the flat-field frames and the sky
background variations. Bad pixels (as identified by Elixir), cosmic rays
(detected using the Laplacian filter of \citealt{vanDokkum01}),
satellite trails, and saturated areas are set to zero in the weight
maps. An object catalog is then produced using
SExtractor~\citep{Sex}, and 
point-like objects are used to derive
an image quality (IQ) estimate.
The sky background map computed by
SExtractor is then subtracted from the image. 
We additionally perform
aperture photometry on the objects of the SExtractor catalog for the purpose of
photometric calibration (see Section~\ref{section:photometric_calibration}).

\subsection{Measurement of supernova fluxes}
\label{sec:meas-supern-flux}

For each supernova candidate, the image with the best IQ (subsequently 
called ``reference'') is identified, and all other images (both science images
and their weight maps) are resampled to the pixel grid defined by this
reference. 
The variations of the Jacobian of the geometrical
transformations, which translate into photometric non-uniformities in
the re-sampled images, are sufficiently small (below the millimag level)
to be ignored.  
We then derive the convolution kernels that would
match the PSF (modeled using the DAOPHOT package \citealt{DAOPHOT}) 
of the reference image to the PSF of the other resampled
science images, {\it but we do not perform the convolutions}. These
convolution kernels not only match the PSFs, but also contain the
photometric ratios of each image to the reference. 
We ensure that these photometric ratios are spatially uniform by 
imposing 
a spatially uniform kernel integral, but allow for spatial kernel shape
variations as the images may have spatially varying PSFs.
Following ~\cite{Alard00}, the kernel is fit on several hundred
objects selected for their high, though unsaturated, peak flux. The
kernel fit is made more robust by excluding objects with large residuals
and iterating. 

Our approach to the differential flux measurement of a SN is
to simultaneously fit all images in a given filter
with a model that includes (i) a spatially variable galaxy (constant
with time), and (ii) a time-variable point source (the supernova). 
The model is described in detail in~\cite{Fabbro01}. 
The shape of the galaxy and positions of both
galaxy and supernova are fit globally. The intensity $D_{i,p}$ in a
pixel $p$ of image $i$ is modeled as:
\begin{equation}
\centering
D_{i,p} = \left [ (f_i P_{ref} + g )\otimes k_i \right ]_p + b_i 
\label{eq:photom_model}
\end{equation}
where $f_i$ are the supernova fluxes, $P_{ref}$ is the PSF of the
reference image centered on the SN position; $k_i$ is the convolution
kernel that matches the PSF of the reference image to the PSF of image
$i$; $g$ is the intensity of the host galaxy in the reference image, 
and $b_i$ is a local (sky)
background in image $i$.  The non parametric galaxy ``model'' $g$ is made of
independent pixels which represent the galaxy in the best IQ image. 
All fluxes ($f_i$) are expressed in units of the reference image flux.

The fit parameters are: the supernova position
and the galaxy pixel values (common to all images), the supernova
fluxes, and a constant sky background (different for each
image). In some images in the series, the supernova flux is known to
be absent or negligible; these frames enter the fit as ``zero flux
images'' and are thus used to determine the values of the galaxy pixels.
The least-squares photometric fit minimizes:
\begin{equation}
\centering
\chi^2 = \sum_{i,p}  W_{i,p}\ (D_{i,p} - I_{i,p})^2
\label{eq:chi2_photo}
\end{equation}
where $I_{i,p}$ and $ W_{i,p}$ are the image and weight values of
pixel $p$ in image $i$, and the sums run over all images that contain
the SN position, and all pixels in the fitted stamp of this image.

Note that this method does not involve any real image convolution: the
fitted model possesses the PSF of the reference image, and it is the model
that is convolved to
match the PSF of every other image.
We typically fit 50x50 galaxy pixels and several hundred images, and
each SN fit usually has 2000 to 3000 parameters. The fit is run once,
5$\sigma$ outlier pixels are removed, and the fit is run again.

The photometric fit yields values of the fit parameters along
with a covariance matrix. There are obvious correlations between SN
fluxes and galaxy brightness, between these two parameters and the
background level, and between the SN position and
the flux, for any given image. More importantly, the uncertainty in the
SN position and the galaxy brightness introduces correlations between
fluxes at different epochs that have to be taken into account when
analyzing the light-curves. Note that flux variances and the correlations 
between 
fluxes decrease when adding more ``zero flux images'' into the fit.  
It will therefore be possible to derive an improved
photometry for most of the events presented in this paper,
when the fields are observed again and more
images without SN light are available.

\subsection{Flux uncertainties}

Once the photometric fit has converged, the parameter covariance
matrix (including flux variances and covariances) is derived. 
This Section addresses the accuracy of these uncertainties,
in particular the flux variances and covariances, which are used as
inputs to the subsequent light-curve fit.  

The normalization of the
parameter covariance matrix directly reflects the normalization of
image weights. We checked that the weights are on average properly
normalized because the minimum $\chi^2$ per degree of freedom is very
close to 1 (we find 1.05 on average).  However, this does not imply
mathematically that the flux uncertainties are properly normalized,
because equation (\ref{eq:chi2_photo}) neglects the correlations
between neighboring pixels introduced by image re-sampling. We
considered accounting for these correlations; however, this would make
the fitting code intolerably slow, as the resulting $\chi^2$ would be
non-diagonal.  
Using approximate errors in least squares (such as ignoring
correlations) increases the actual variance of the estimators, but in
the case considered here, the loss in photometric accuracy is below 1\%.
The real drawback of ignoring pixel correlations is that
parameter uncertainties extracted from the fit are underestimated
(since pixel correlations are positive); this is a product of any photometry
method that assumes uncorrelated pixels on re-sampled or convolved
images.  Our geometric alignment technique, used to align images prior
to the flux measurement as described in
Section~\ref{sec:meas-supern-flux}, uses a 3x3 pixel quadratic
re-sampling kernel,
which 
produces output pixels with an average variance of 80\% of the input
pixel variance, where the remaining 20\% contributes to covariance in nearby pixels.
We checked that flux variances (and covariances)
computed assuming independent pixels are also underestimated by the
same amount: on average, a 25\% increase is required.

In order to derive accurate
uncertainties, we used the fact that for
each epoch, several images are available which measure the same
object flux.  Estimating fluxes on individual exposures rather than on
stacks per night preserves the photometric precision since a common
position is fit using all images. It also allows a check on the
consistency of fluxes measured within a night. We therefore fit a
common flux per night to the fluxes measured on each individual image
by minimizing a $\chi^2_{n}$ (where $n$ stands for nights); this matrix is
non-diagonal because the differential photometry produces correlated
fluxes.  The $\chi^2_{n}$ contribution of every individual image is
evaluated, and outliers $>5\sigma$  (due to, for
example, unidentified cosmic rays) are discarded; this cut
eliminates 1.4\% of the measurements on average.  The covariance of
the per-night fluxes is then extracted, and normalized so that the minimum
$\chi^2_n$ per degree of freedom is 1. 
This translates into an ``effective'' flux uncertainty derived from
the scatter of repeated observations rather than from first
principles. If the only source of noise (beyond photon statistics) were
pixel correlations introduced by image resampling, we would expect an
average $\chi^2_{n}/N_{dof}$ of 1.25, as all flux variances are on
average under-estimated by 25\%. Our average value is 1.55; hence we
conclude that our photometric uncertainties are only $\sim12$\%
($\sqrt(1.55/1.25)-1$) larger than photon statistics, 
leaving little margin for drastic improvement.

\begin{figure}
\centering
\includegraphics[width=\linewidth]{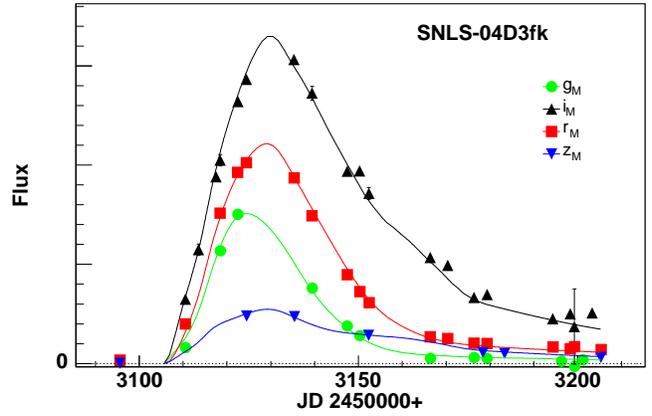}
\caption{Observed light-curves points 
of the SN Ia SNLS-04D3fk in $\gme$, $\rme$, $\ime$ and $\zme$ bands, 
along with the multi-color light-curve model (described
in Section \ref{section:lightcurve_fit_model}).
Note the regular sampling of the observations both before and after 
maximum light. 
With a SN redshift of 0.358, the four measured pass-bands 
lie in the wavelength range of the light-curve model, 
defined by rest-frame $U$ to $R$ bands, and 
all light-curves points are therefore fitted simultaneously with only 
four free parameters (photometric
normalization, date of maximum, a stretch and a color parameter). 
\label{figure:intermediate_lc}
}
\end{figure}

\begin{figure}
\centering
\includegraphics[width=\linewidth]{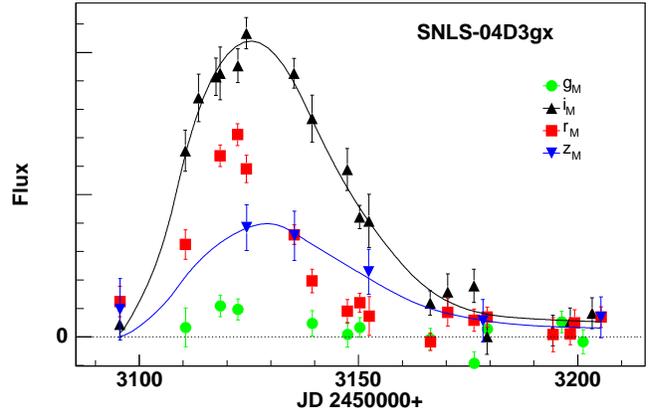}
\caption{Observed light-curves points of the SN Ia SNLS-04D3gx at z=0.91.  
With a SN redshift of 0.91, only two of the measured pass-bands 
lie in the wavelength range of the light-curve model, 
defined by rest-frame $U$ to $R$ bands, and are therefore used in the fit (shown 
as solid lines). 
Note the excellent quality of the
photometry at this high redshift value. Note
also the clear signal observed in $\rme$ and even in $\gme$, 
which correspond to central wavelength
of respectively $\lambda\sim3200\AA$ and $\lambda\sim2500\AA$ in the SN 
rest-frame. 
\label{fig:04D3gx}}
\end{figure}

Table \ref{table:fitnight_summary}
summarizes the statistics of the differential photometry fits in each
filter. The larger values of $\chi^2_n/N_{dof}$ in $\ime$ and $\zme$
probably indicate contributions from residual fringes. Examples of SNe~Ia
light-curves points are presented in 
Figures~\ref{figure:intermediate_lc} and~\ref{fig:04D3gx}
showing SNe at $z=0.358$ and $z=0.91$ respectively. 
Also shown on these Figures are the results of the light-curves fits 
described in Section~\ref{section:lightcurve_fit_model}.

\begin{table}
\centering
\begin{tabular}{c||c|c|c|c}
Band   & Average nb. & Average nb. & $\chi^2_n$ & Central\\
       & of images      &  of epochs          &  per d. o. f. & wavelength \\
\hline
 $\gme$ & 40 & 9.8 & 1.50    & 4860\\
 $\rme$ & 75 & 14.4 &  1.40  & 6227\\
 $\ime$ & 100 & 14.8 &  1.63 & 7618\\
 $\zme$ & 60 & 7.9 &  1.70   & 8823\\ 
\end{tabular}
\caption{
\label{table:fitnight_summary}
Average number of images and nights per band for 
each SNLS light-curve. Note that there is less data in $\gme$ and $\zme$.
The $\chi^2_n$ column refers to the last fit that imposes equal
fluxes on a given night. The expected value is 1.25 (due to pixel correlations),
so we face a moderate scatter excess of about 12\% over photon statistics. 
The larger values in $\ime$ and $\zme$ indicate that fringes play a role in this excess. The last column displays the average wavelength of the effective
filters in~\AA}
\end{table}

The next Section discusses how accurately the SN fluxes can be
extracted from the science frames relative to nearby field stars, i.e.
how well the method assigns magnitudes to SNe, given magnitudes of the
field stars which are used for photometric calibration, called
tertiary standards hereafter.

\subsection{Photometric alignment of supernovae relative to tertiary standards}
\label{section:calibration_of_sn_fluxes}

The SN flux measurement technique of
Section~\ref{sec:meas-supern-flux} delivers SN fluxes on the same
photometric scale as the reference image. In this Section, we discuss
how we measure ratios of the SN fluxes to those of the tertiary
standards (namely stars in the SNLS fields).
The absolute flux calibration of the tertiary standards themselves is 
discussed in Section~\ref{section:photometric_calibration}.

The image model that we use to measure the SN fluxes (eq.
\ref{eq:photom_model}) can also be adapted to fit the tertiary standards
by setting the ``underlying galaxy model'' to zero.  We
measure the fluxes of field stars by running the same simultaneous fit
to the images used for the supernovae, but without the ``zero-flux''
images, and without an underlying galaxy.  As this fitting technique
matches that used for the SNe as closely as possible, most of the
systematics involved (such as astrometric alignment residuals, PSF
model uncertainties, and the convolution kernel modeling) 
cancel in the flux ratios.

For each tertiary standard (around 50 per CCD), we obtain one flux for
each image (as done for the SNe), expressed in the same units. From the
magnitudes of these fitted stars, we can extract a photometric zero
point for the PSF photometry for every star on every image, 
which should be identical within measurement
uncertainties. Several systematic checks were performed to search for
trends in the fitted zero-points as a function of several variables
(including image number, star magnitude, and 
star color); no significant trends were detected.
As zero-points are obtained from single measurements on single
images, the individual measurements are both numerous and
noisy, with a typical r.m.s of 0.03 mag; however since they have the
same expectation value, we averaged them using a robust
fit to the distribution peak to obtain a single zero-point per observed
filter.

To test how accurately the ratio of SN flux to tertiary standard stars is
retrieved by our technique, we tested the method on simulated SNe.
For each artificial supernova, we selected a random host galaxy, a
neighboring bright star (the model star), and a down-scale ratio ($r$). For
half of the images that enter the fit, we superimposed a
scaled-down copy (by a factor $r$) of the model on the host galaxy. 
We rounded the artificial position at an integer pixel offset from the 
model star to avoid re-sampling.
We then performed the full SN fit (i.e. one that allows for an
underlying galaxy model and ``zero flux images'') at the position of
the artificial object, and performed the calibration star fit (i.e. one with
no galaxy mode and no ``zero-flux images'') at the original position
of the model star. This matches exactly the technique used for the
measurement and calibration of a real SN.  We then compared the
recovered flux ratio to the (known) down-scale ratio.  

We found no significant bias as a function of SN flux or galaxy
brightness at the level of 1\%, except at signal-to-noise (S/N) ratios
(integrated over the whole light-curve) below 10.  At a
S/N ratio of 10, fluxes are on average
underestimated by less than 1\%; this bias rises to about 3\% at a S/N
ratio of 7. This small flux bias disappears when the fitted object position
is fixed, as expected because the fit is then linear.  For this
reason, when fitting $\zme$ light-curves of objects at
$z>0.7$, for which the S/N is expected to be low, we use the fixed SN
position from that obtained from the $\ime$ and $\rme$ fits. 

Given the statistics of our simulations, the systematic uncertainty of
SN fluxes due to the photometric method employed is less than 1\%
across the range of S/N we encounter in real data, and the observed
scatter of the retrieved ``fake SNe'' fluxes behaves in the same way
as that for real SNe.  Over a limited range of S/N (more than 100
integrated over the whole light-curve), we can exclude biases at the
0.002 mag level. Our upper limits for a flux bias have a
negligible impact on the cosmological conclusions drawn from the
sample described here, and will likely be improved with further
detailed simulations. 

\section{Photometric calibration}
\label{section:photometric_calibration}

The supernova light-curves produced by the techniques described in
Section~\ref{sec:meas-supern-flux} are calibrated relative to nearby
field stars (the tertiary standards). Our next step is to place these
instrumental fluxes onto a photometric calibration system using
observations of stars of known magnitudes.

\subsection{Photometric calibration of tertiary standards}
\label{section:calibration_of_field_stars}

Several standard star calibration catalogs are available in the
literature, such as the \citet{Landolt83,Landolt92} Johnson-Cousins
(Vega-based) $UBVRI$ system, or the \citet{smith02} $u'g'r'i'z'$
AB-magnitude system which is used to calibrate the
Sloan Digital Sky Survey (SDSS).  However, there are systematic errors
affecting the transformations between the \citet{smith02} system and the widely used
Landolt system. As discussed in \cite{Fukugita96}, these errors arise
from various sources, for example uncertainties in the
cross-calibration of the spectral energy distributions of the AB
fundamental standard stars relative to that of Vega. Since the nearby
SNe used in our cosmological fits were extracted from the literature
and are typically calibrated using the standard star catalogs of
\citet{Landolt92}, we adopted the same calibration source for our
high-redshift sample.  This avoids introducing additional systematic
uncertainties between the distant and nearby SN fluxes, which
are used to determine the cosmological parameters.
To eliminate uncertainties associated with color corrections, we
derive magnitudes in the natural MegaCam filter system.  

Both standard and science fields were repeatedly observed over a
period of about 18 months.  Photometric nights were selected using the
CFHT ``Skyprobe'' instrument \citep{Skyprobe}, which monitors
atmospheric transparency in the direction that the telescope is pointing. Only
the 50\% of nights with the smallest scatter in transparency were
considered. For each night, stars were selected in the science fields
and  their aperture fluxes  measured and corrected to an
airmass of 1 using the average atmospheric extinction of Mauna Kea.
These aperture fluxes were then averaged, allowing for photometric
ratios between exposures.  Stable observing conditions were indicated
by a very small scatter in these photometric ratios (typically
0.2\%); again the averaging was robust, with 5-$\sigma$ deviations
rejected.  Observations of the Landolt standard star fields were
processed in the same manner, though their fluxes were not averaged.
The apertures were chosen sufficiently large (about 6\arcsec\ in
diameter) to bring the variations of aperture corrections across the
mosaic below 0.005 mag. However, since fluxes are measured in the same
way and in the same apertures in science images and standard star
fields, we did not apply any aperture correction.

Using standard star observations, we first determined zero-points 
by fitting linear color transformations and zero-points to each night
and filter,  however with color slopes common to all nights.
In order to account for possible non-linearities in the 
Landolt to MegaCam color relations,
the observed color-color relations were then compared to
synthetic ones derived from spectrophotometric standards. This led
to shifts of roughly 0.01 in all bands other than $\gme$, for which the 
shift was 0.03 due to the nontrivial relation to $B$ and $V$.

\begin{figure}
\centering
\includegraphics[width=\linewidth]{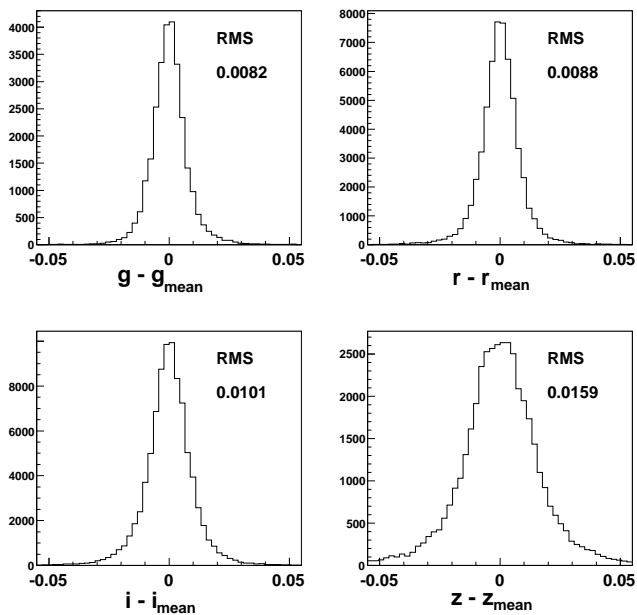}
\caption{The calibration residuals --- {\em i.e.} the residuals around 
the mean 
magnitude of each Deep field tertiary 
standard--- in the bands $\gme$, $\rme$, $\ime$ and $\zme$,
for all CCDs and fields, with one entry per star and epoch. 
The dispersion is below 1\% in $\gme$, $\rme$ and $\ime$,
and about 1.5\% in $\zme$.
\label{figure:calibration_residuals}}
\end{figure}

We then applied the zero-points appropriate for each night to the
catalog of science field stars of that same night. These magnitudes were 
averaged
robustly, rejecting 5-$\sigma$ outliers, and the average standard star
observations were merged.
Figure \ref{figure:calibration_residuals} shows the dispersion of the
calibration residuals in the $\gme$, $\rme$, $\ime$ and $\zme$ bands.
The observed standard deviation, which sets the upper bound to the
repeatability of the photometric measurements, is about or below 0.01~mag in
$\gme$, $\rme$ and $\ime$, and about 0.016~mag in $\zme$.

For each of the four SNLS fields, a catalog of tertiary standards was
produced using the procedure described above.  These catalogs were
then used to calibrate the supernova fluxes, as described in 
Section~\ref{section:calibration_of_sn_fluxes}.  The dominant uncertainty in
the photometric scale of these catalogs comes from the determination
of the color-color relations of the standard star measurements. For
the $\gme$, $\rme$ and $\ime$ bands, a zero-point offset of 0.01 mag
would easily be detected; hence we took this value as a conservative
uncertainty estimate. The $\zme$ band is affected by a larger
measurement noise, and it is calibrated with respect to $I$ and $R-I$
Landolt measurements.  We therefore attributed to it a larger zero
point uncertainty of 0.03 mag.

The MegaCam shutter is designed to preserve the mosaic
illumination uniformity. Nevertheless,  
the shutter precision is a potential source of systematic
uncertainties, given (1) the possible non uniformities due to the
shutter motion and (2) the exposure time differences between the
calibration images (a few seconds) and the science images (hundreds of
seconds). For MegaCam, the {\em actual} exposure time
is measured and reported for each exposure, using dedicated
sensors. The shutter precision was investigated by
\citet{Cuillandre05} and it was shown that the non-uniformity due to the
shutter is less than 0.3\% across the mosaic. Short and long exposures
of the same fields were also compared. The systematic flux differences
between the exposures were found to be below 1\% (r.m.s).

\subsection{The MegaCam and Landolt instrumental filters}
\label{section:instrumental_filters}

As the supernova fluxes are measured in the instrumental filter
system, the MegaCam transmission functions (up to an arbitrary constant) 
are needed in order to
correctly interpret the SN photometry.  Similarly, for the published
nearby supernovae which are reported in Landolt magnitudes, the filter
responses of the Landolt system are required.

For the MegaCam filters, we used the measurements provided by the
manufacturer, multiplied by the CCD quantum efficiency, the
MegaPrime wide-field corrector transmission function, the CFHT primary
mirror reflectivity, and the average atmospheric transmission at Mauna
Kea. As an additional check, we computed synthetic
MegaCam-SDSS color terms using the synthetic transmissions of the
SDSS 2.5-m telescope \citep{SkyServerInstrument} and spectrophotometric
standards taken from \citet{Pickles98} and \citet{Gunn83}. Since the SDSS science catalog
\citep{Finkbeiner94,Raddick02,SkyServer} shares thousands of objects
with two of the four fields repeatedly observed with MegaCam, we were
able to compare these synthetic color transformations with the
observed transformations. We found a good agreement, with uncertainties 
at the 1\% level.  This constrains the central wavelengths of the
MegaCam band passes to within 10 to 15~\AA\ with respect to the SDSS 2.5m
band passes.

The choice of filter band passes to use for Landolt-based observations is not
unique.  Most previous supernova cosmology works assumed that the
determinations of \citet{Bessel90} describe the effective Landolt
system well, although the author himself questions this fact,
explicitly warning that the Landolt system {\em ``is
  not a good match to the standard system''} -- {\em i.e.} the historical
Johnsons-Cousins system.
Fortunately, \cite{Hamuy92, Hamuy94} provide spectrophotometric
measurements of a few objects measured in \cite{Landolt92b}; this
enabled us to compare synthetic magnitudes computed using Bessell
transmissions with Landolt measurements of the same objects. This
comparison reveals small residual color terms which vanish if the $B$,
$V$, $R$ and $I$ Bessell filters are blue-shifted by 41, 27, 21 and 25~\AA\ 
respectively. Furthermore, if one were to assume
that the Bessell filters describe
the Landolt system, this would lead to synthetic MegaCam-Landolt color terms
significantly different from the measured ones; the blue
shifts determined above bring them into excellent agreement. We 
therefore assumed
that the Landolt catalog magnitudes refer to blue-shifted Bessell
filters, with a typical central wavelength uncertainty of 10 to 15~\AA, 
corresponding roughly to a 0.01 accuracy for the color terms.

\subsection{Converting magnitudes to fluxes}
\label{section:converting_magnitudes_to_fluxes}

Given the variations with time of the cosmological scale factor
$a(t)$, one can predict the evolution with redshift of the observed
flux of classes of objects of reproducible luminosity though not necessarily known. This is why the cosmological conclusions that
can be drawn from flux measurements rely on flux ratios of distant to
nearby SNe, preferably measured in similar rest-frame pass-bands.  The
measured SNe magnitudes must therefore be converted to fluxes at
some point in the analysis. 

The flux in an imaginary rest-frame band of
transmission $T_{rest}$ for a SN at redshift $z$ is deduced from the
magnitude $m(T_{obs})$ measured in an observer band of transmission
$T_{obs}$ via:
\begin{equation}
\begin{split}
f(z,T_{rest})  =  10^{-0.4(m(T_{obs}) - m_{ref}(T_{obs}))} \\
        \times \frac{\int \phi_{SN}(\lambda) T_{rest}(\lambda) d\lambda}
             {\int \phi_{SN}(\lambda) T_{obs}(\lambda(1+z)) d\lambda}
        \int \phi_{ref}(\lambda) T_{obs} (\lambda) d\lambda
\label{eqn:f_z_trest}
\end{split}
\end{equation}
where $\phi_{SN}$ is the spectrum of the SN, $m_{ref}(T)$ is the
magnitude of some reference star that was used as a calibrator, and
$\phi_{ref}$ is its spectrum.  In this expression, the product of the
first and third terms gives the integrated flux in the observed band,
and the second term scales this integrated flux to the rest-frame band.
We measure only $m(T_{obs})-m_{ref}(T_{obs})$ (if the reference
star is directly observed), or only $m(T_{obs})$  (if a non-observed star
-- e.g. Vega -- is used as the reference). The reference spectrum, $\phi_{ref}$,
must be taken from the literature, as well as $m_{ref}(T_{obs})$ if the 
reference is not directly observed. The supernova spectrum, $\phi_{SN}$, is taken
to be a template spectrum appropriately warped to
reproduce the observed color of the SN (as described in \citealt{Guy05}).
The quantity $f(z,T_{rest})$
scales as the inverse square of a luminosity distance:
\begin{equation}
\centering
\frac{f(z_1, T_{rest})}{f(z_2, T_{rest})} = \left( \frac{d_L(z_2)}{d_L(z_1)} \right)^2
\label{eqn:flux_luminosity_distance}
\end{equation}

This conversion of a measured magnitude to a
rest-frame flux (or a rest-frame magnitude) is usually integrated in the
so-called cross-filter k-corrections \citep{Kim96,Nugent02}. In our case,
it is integrated in the light-curve fit \citep{Guy05}.
(See \cite{Guy05} for a discussion of the precise definitions
of spectra and transmissions that enter into $f(z,T_{rest})$.)

Inspecting eq. \ref{eqn:f_z_trest}, we  first note that the normalizations 
of $T_{obs}$ and $\phi_{SN}$ cancel.
The width of $T_{obs}$ is a second order effect.
When forming the ratio of two such quantities for two different SN, 
the normalization of $\phi_{ref}$ does not matter, nor the normalization
of $T_{rest}$, provided the same $T_{rest}$ is chosen for both objects. The width of
 $T_{rest}$  matters only at the second order. The factors
that do enter as first order effects are:
\begin{itemize}
\item $\int \phi_{ref}(\lambda) T_{obs,1}(\lambda(1+z_1)) d\lambda
 / \int  \phi_{ref}(\lambda) T_{obs,2}(\lambda(1+z_2)) d\lambda $ ,
    which requires both the spectrum of a reference
   and the band passes of the observing systems, i.e. to first order, 
their central wavelengths,
\item $m_{ref}(T_{obs,1}) - m_{ref}(T_{obs,2})$ , i.e. the color of the 
reference. When comparing distant and nearby SNe, we typically rely on 
$B-R$ or $B-I$ colors, 
\item and obviously, the SNe measured magnitudes, or, more precisely,
their difference.
\end{itemize}

We choose to use Vega as the reference star. An accurate spectrum of
Vega was assembled by \cite{Hayes85}. Some subtle differences are
found by a more recent HST measurement \citep{Bohlin04} but they only
marginally affect broadband photometry: differences within the 1\%
uncertainty quoted in \cite{Hayes85} are found and we will assign this
uncertainty to the Vega broadband fluxes.  We use the HST-based
measurement because it extends into the UV and NIR and hence is safe for
the blue side of the $U$ band and in the $\zme$ band.  For Vega,we adopt
the magnitudes ($U$,$B$,$V$,$R_c$,$I_c$) = (0.02, 0.03, 0.03, 0.03, 0.024) (\cite
{Fukugita96} and references therein). For other bands,
a simple interpolation is adequate. Note that only Vega colors impact on
cosmological measurements.

A possible shortcut consists in relying on spectrophotometric
standards \citep{Hamuy92,Hamuy94} which
also have magnitudes on the Landolt system \citep{Landolt92b}. When
we compare synthetic Vega magnitudes of these objects with the
photometric measurements, we find excellent matching of colors (at
better than the 1\% level), indicating that choosing Vega or 
spectrophotometric fluxes as the reference makes little practical difference.

\subsection{Photometric calibration summary}

We constructed catalogs of tertiary standard stars in the SNLS fields,
expressed in MegaCam natural magnitudes, and defined on the Landolt
standard system.  The repeatability of measurements of a single star
on a given epoch (including measurement noise) is about or below 0.01~mag
r.m.s in $\gme$, $\rme$ and $\ime$, and about 0.016~mag in $\zme$.  From
standard star observations, we set conservative uncertainties of the
overall scales of 0.01 mag in $\gme$, $\rme$ and $\ime$ and 0.03 in
$\zme$.  The MegaCam central wavelengths are constrained by color
terms with respect to both the SDSS 2.5m telescope and the Landolt 
catalog to
within 10 to 15~\AA.  The central wavelengths of the band passes of
the Landolt catalog are found slightly offset with respect to
\cite{Bessel90}, using spectrophotometric measurements of a subsample
of this catalog.

\section{Light-curve fit and cosmological analysis}
\label{section:snIa_lc_cosmo}

To derive the brightness, light-curve shape and SN color estimates
required for the cosmological analysis, the time sequence of photometric
measurements for each SN was fit using a SN light-curve model.  This procedure
is discussed in this section together with 
the nearby and distant SN Ia samples selection and the cosmological analysis.

\subsection{The SN~Ia light-curve model}
\label{section:lightcurve_fit_model}

We fit the SN~Ia light-curves in two or more bands using the
SALT light-curve model (\citealt{Guy05}) which returns
the supernova rest-frame $B$-band magnitude $m^*_B$, 
a single shape parameter $s$ and a single color parameter $c$. 
The supernova rest-frame $B$-band magnitude at the date
of its maximum luminosity in $B$ is defined as:
$$
 m^*_B = -2.5\ \log_{10}\left(\frac{f(z, T^*_B, t=t_{max,B}}{(1+z) \int \phi_{ref}(\lambda) T_B(\lambda) d\lambda}\right)
$$
where $T^*_B(\lambda) \equiv T_B(\lambda/(1+z)) \equiv T_{rest}(B)$ is 
the rest-frame
$B$-band transmission, and  $f(z, T^*_B, t=t_{max,B})$ is defined
by eq. \ref{eqn:f_z_trest}. The stretch factor $s$ is
similar to that described in \cite{Perlmutter97}: it parameterizes
the brighter-slower relation, originally described in \cite{Phillips93}, 
by stretching the time axis of a unique light-curve template; 
$s=1$ is defined in 
rest-frame $B$ for the time interval
$-15$ to $+35$ days using the \citet{Goldhaber01} $B$-band template.
For bands other than B, stretch is a parameter that indexes light-curve 
shape variability.
The rest-frame color $c$ is defined by $c = (B-V)_{B\,max} +
0.057$: it is a color excess (or deficit) with respect to a
fiducial SN~Ia (for which $B-V=-0.057$ at $B$-band maximum light).
Note that
the color $c$ is not just a measure of host galaxy extinction: $c$ 
can accommodate both reddening by dust and any intrinsic color effect
dependent or not on $s$.
The reference value ($-0.057$)
can be changed without changing the cosmological conclusions, 
given the distance estimator we use (see
Section~\ref{section:cosmo-fit}).

The light-curve model was trained on very nearby supernovae 
(mostly at $z<0.015$) published
in the literature (see \citealt{Guy05} for the selection of these
objects).
Note that these training objects were {\it not} used
in the Hubble diagram described in this paper.
The SALT light-curve model
generates light-curves in the observed bands at a given redshift, 
SALT also incorporates corrections for the Milky Way extinction, 
using the dust maps of \citet{Schlegel98} coupled with
the extinction law of \citet{Cardelli89}.
The rest-frame coverage of SALT extends from 3460 to 6500~\AA\ (i.e. slightly 
bluewards from $U$ to $R$).
We require that photometry is available in at least 2 measured
bands with central wavelengths within this wavelength range to 
consider a SN for the
cosmological analysis.  Light curves in the $\zme$
band become essential for $z>0.80$, since at these redshifts, $\rme$
corresponds to rest-frame $\lambda < 3460~\AA$.
All observed bands are fitted simultaneously,
with common stretch and color parameters, 
global intensity and date of $B$-band maximum light.  
Making use of $U$-, $B$- and $V$-band measurements of nearby SNe~Ia
from the literature (mostly from \citealt{Hamuy96b,Riess99a,Jha02}), 
\cite{Guy05} have constructed
a distance estimator using either $U$- and $B$-band data or $B$- and $V$-band
which shows a dispersion of 0.16 mag
around the Hubble line.
The fitted global intensity is then translated into a rest-frame-$B$ observed
magnitude at maximum light ($m_B^*$)
which does not include any correction for brighter-slower or
brighter-bluer relations.  

The light-curve fit is carried out in two steps. The first fit uses
all photometric data points to obtain a date of maximum light in the
$B$-band. All points outside the range [$-15,+35$] rest-frame days from
maximum are then rejected, and the data refit.  This restriction
avoids the dangers of comparing light-curve parameters derived from
data with different phase coverage: nearby SNe usually have
photometric data after maximum light, but not always before maximum
when the SN is rising, and almost never before $-15$ days. By
contrast, SNLS objects have photometric sampling that is essentially
independent of the phase of the light-curve because of the
rolling-search observing mode, though late-time data (in the
exponential tail) often has a poor S/N, or is absent due to field
visibility.

\subsection{The SN~Ia samples}
\label{section:snIa-sample}

The cosmological analysis requires assembling a sample of nearby
and distant SNe Ia. 

We assembled a nearby SN~Ia sample from the literature.  
Events with redshifts below $z=0.015$ were rejected 
to limit the influence of peculiar velocities.
We further retained only objects whose
first photometric point was no more than 5 days after maximum light.  
To check for possible biases that this latter procedure might have
introduced, we fitted subsets of data from objects with pre-maximum photometry.
Our distance estimator (see Section~\ref{section:cosmo-fit}) was 
found to be unaffected if the data started up to 7 days after maximum
light.
A sample of 44 nearby SNe~Ia matched our requirements.
Table~\ref{table:nearbydata} gives the SN name, redshift and filters
used in the light-curve fits, as well as fitted rest-frame
$B$-band magnitude and values of the parameters $s$ and $c$.

For this paper, we considered only distant SNe Ia 
that were discovered and 
followed during the first year of SNLS since this data set already  
constitutes the largest well controlled homogeneous sample of distant SN Ia. 
As discussed in Section~\ref{section:first_year}, 91 objects were
spectroscopically identified as ``Ia'' or ``Ia*'', with a date of
maximum light before July 15, 2004. Ten of these are not
yet analyzed: 5 because images uncontaminated by SN light were
not available at the time of this analysis, and 5 due to a limitation 
of our reduction pipeline which does not yet handle field regions 
observed with different CCDs.
Six SNe have incomplete data due to either instrument failures, 
or persistent bad weather and two SNe, SNLS-03D3bb and SNLS-03D4cj,
which happen to be spectroscopically peculiar 
(see \citeauthor{Ellis05}, in prep.) have 
photometric data incompatible with the light-curve model.

The resulting fit parameters of the remaining
73 ``Ia''+''Ia*'' SNe are given
in Table~\ref{table:snlsdata} and 
examples of light-curves measured in the four
MegaCam bands are shown in Figures~\ref{figure:intermediate_lc}
and \ref{fig:04D3gx}, together with the result of the light-curve fit.

\subsection{Host galaxy extinction}

There is no consensus on how to correct for host galaxy extinction
affecting high redshift SNe~Ia.  The pioneering SN cosmology papers
(\citealt{Riess98b}, \citealt{Perlmutter99}) typically observed in only one
or two filters, and so had  little or no color information
with which to perform  extinction corrections.  Subsequent
papers either selected low-extinction subsamples based on host galaxy
diagnostics \citep{Sullivan03}, or used multicolor information
together with an assumed color of an unreddened SN to make extinction
corrections on a subset of the data \citep{Knop03,Tonry03}.

These techniques have their drawbacks: the intrinsic color of
SNe Ia has some dispersion, and measured colors often have large
statistical errors in high-redshift data sets. When these two color
uncertainties are multiplied by the ratio of total to selective
absorption, $R_B\simeq 4$, the resulting error can be very large. 
To circumvent this, some studies
used Bayesian priors 
(e.g.  \citealt{Riess98b,Tonry03,Riess04,Barris04}).  Other authors
argue that this biases the results (e.g. \citealt{Perlmutter99,Knop03}).

Here we employ a technique that makes use of color information to
empirically improve distance estimates to SNe Ia.
We exploit the fact that the SN color acts in the same
direction as reddening due to dust -- i.e. redder SNe are
intrinsically dimmer, brighter SNe are intrinsically bluer
\citep{Tripp99}.  
By treating the correction between color and brightness empirically,
we avoid model-dependent assumptions 
that can both
artificially inflate the errors and potentially lead to biases in the
determination of cosmological parameters.  Because we have more than
one well-measured color for several SNe, we can perform consistency checks on
this technique -- distances from multiple colors should, and do, 
agree to a remarkable degree of 
precision (Section~\ref{section:SNcolorcompatibility}).

\subsection{Cosmological fits}
\label{section:cosmo-fit}

From the fits to the light-curves 
(Section~\ref{section:lightcurve_fit_model}), 
we computed a rest-frame-$B$
magnitude, which, for perfect standard candles, should vary with
redshift according to the luminosity distance. This rest-frame-$B$
magnitude refers to {\it observed} brightness, and therefore does 
not account for
brighter-slower and brighter-bluer correlations (see~\citealt{Guy05} and
references therein).  As a distance estimator, we use:
$$
\mu_B = m^{*}_B - M +\alpha(s-1) - \beta c
$$
where $m^{*}_B$, $s$ and $c$ are derived from the fit to the light curves, and
$\alpha$, $\beta$ and the absolute magnitude $M$ are parameters which 
are fitted by minimizing the residuals in the Hubble diagram.
The cosmological fit is actually performed by minimizing:
$$
\chi^2 = \sum_{objects} \frac{\left ( \mu_B - 5 \log_{10} (d_L(\theta,z)/ 10pc)\right)^2}{\sigma^2(\mu_B)+\sigma^2_{int}} ,
$$
where $\theta$ stands for the cosmological parameters that define
the fitted model (with the exception of $H_0$), 
$d_L$ is the luminosity distance,
and $\sigma_{int}$ is the intrinsic dispersion of SN absolute magnitudes.
We minimize with respect to $\theta$, $\alpha$, $\beta$ and $M$.
Since $d_L$ scales as $1/H_0$, only $M$ depends on $H_0$.
The definition of $\sigma^2(\mu_B)$, the measurement variance,
requires some care. First, one has to account for the
full covariance matrix of $m^{*}_B$, $s$ and $c$ from the light-curve
fit. Second, $\sigma(\mu_B)$ depends on $\alpha$ and $\beta$; minimizing with
respect to them introduces a bias towards increasing errors in order to
decrease the $\chi^2$, as originally noted in \cite{Tripp98}. When
minimizing, we therefore fix the values of $\alpha$ and $\beta$
entering the uncertainty calculation and update them
iteratively. $\sigma(\mu_B)$ also includes a peculiar velocity contribution
of $300$~km/s. $\sigma_{int}$ is introduced to account for the 
``intrinsic dispersion''
of SNe~Ia. We perform a first fit with an initial value (typically
$0.15$ mag.), and then calculate the $\sigma_{int}$ required to obtain a
reduced $\chi^2=1$. We then refit with this more accurate value. 
We fit 3 cosmologies to the data: a $\Lambda$ cosmology (the
parameters being $\om$ and $\ol$), a flat $\Lambda$ cosmology (with a
single parameter $\om$), and a flat $w$ cosmology, where $w$ is the
constant equation of state of dark energy (the parameters are
$\om$ and $w$). 

\begin{figure}
\begin{center}
\includegraphics[width=\linewidth]{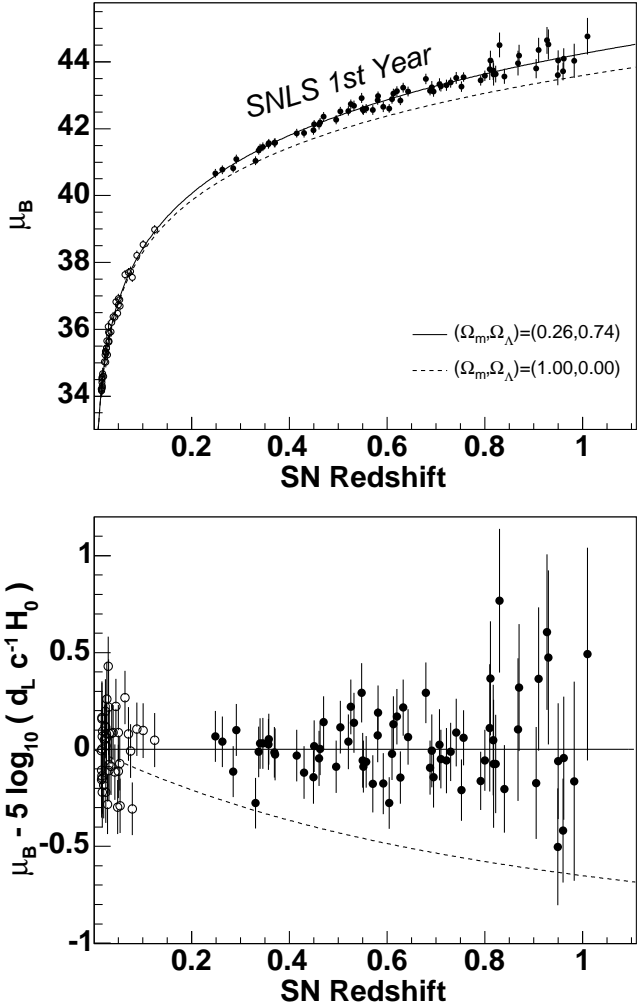}
\caption{Hubble diagram of SNLS and nearby SNe~Ia, with various
cosmologies superimposed. The bottom plot shows the residuals for
the best fit to a flat $\Lambda$ cosmology.
\label{fig:hubble}
}
\end{center}
\end{figure}

The Hubble diagram of SNLS SNe and nearby
data is shown in Figure~\ref{fig:hubble}, together with the best fit
$\Lambda$ cosmology for a flat Universe.
Two events lie more than 3 $\sigma$ away from the Hubble diagram fit:
SNLS-03D4au is 0.5~mag fainter than the best-fit and 
SNLS-03D4bc is 0.8~mag fainter.
Although, keeping or removing these SNe from the fit has a minor effect 
on the final result, they were not kept in the final cosmology fits (since 
they obviously depart from the rest of the population) which 
therefore make use of 44 nearby objects
and 71 SNLS objects. 

The
best-fitting values of $\alpha$ and $\beta$ are $\alpha = 1.52\pm
0.14$ and $\beta=1.57 \pm 0.15$, comparable with previous works using
similar distance estimators (see for example \citealt{Tripp98}).
As discussed by several authors (see \cite{Guy05} and references therein),
the value of
$\beta$ does differ considerably from $R_B=4$, the value expected if
color were only affected by dust reddening.
This discrepancy may be an indicator of intrinsic color variations in the
SN sample (e.g. \citealt{Nobili03}), and/or  variations in $R_B$. For the 
absolute magnitude $M$, we obtain $M=-19.31 \pm0.03 + 5 \log_{10} h_{70}$.

\begin{figure}
\begin{center}
\includegraphics[width=\linewidth]{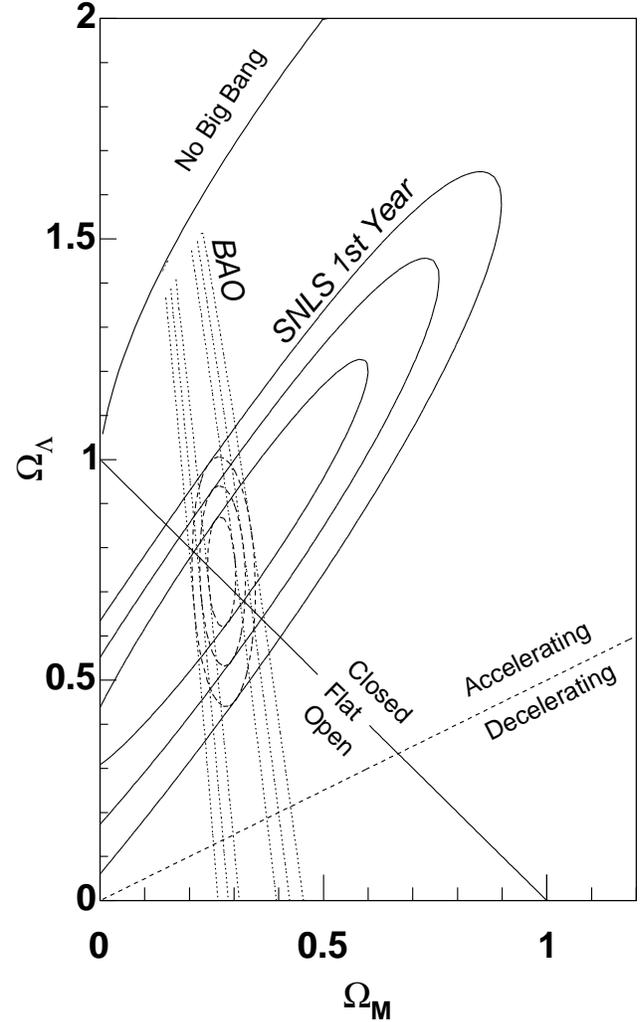}
\caption{Contours at 68.3\%, 95.5\% and 99.7\% confidence levels for 
the fit to an $(\om,\ol)$ cosmology from the SNLS Hubble diagram 
(solid contours),
 the SDSS baryon acoustic oscillations (\citealt{Eisenstein05}, 
dotted lines), and the joint confidence contours (dashed lines).
\label{fig:contours_om_ol}
}
\end{center}
\end{figure}

\begin{figure}
\begin{center}
\includegraphics[width=\linewidth]{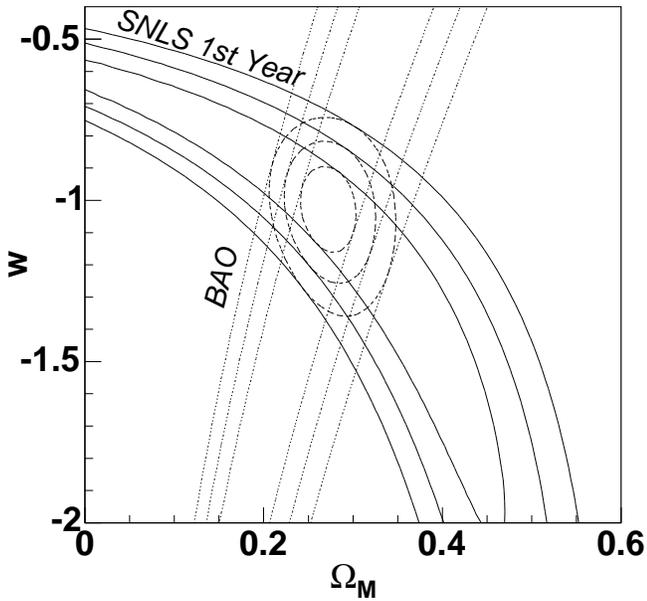}
\caption{Contours at 68.3\%, 95.5\% and 99.7\% confidence levels 
for the fit to a flat $(\om,w)$ cosmology, from the SNLS Hubble diagram 
alone, from the SDSS baryon acoustic oscillations alone \citep{Eisenstein05}, 
and the joint confidence contours.\label{fig:contours_om_w}
}
\end{center}
\end{figure}

The parameters $\alpha$, $\beta$ and $M$ are nuisance
parameters in the cosmological fit, and their
uncertainties must be accounted for in the cosmological error analysis.
The resulting confidence contours are shown in
Figures \ref{fig:contours_om_ol} and \ref{fig:contours_om_w}, together
with the product of these confidence estimates with 
the probability distribution from  baryon acoustic
oscillations (BAO) measured in the SDSS (Eq. 4 in \citealt{Eisenstein05}).
We impose $w=-1$ for the $(\om,\ol)$ contours, and $\ok=0$ for the
$(\om,w)$ contours. Note that the constraints from BAO and SNe~Ia are 
quite complementary. The best-fitting cosmologies are given in 
Table~\ref{tab:cosmo_fits}. 
\begin{center}
\begin{table}[h]
\begin{tabular}{l|c}
fit & parameters (stat only)\\
\hline 
$(\om,\ol)$ &   $(0.31\pm0.21,0.80\pm0.31)$ \\
$(\om-\ol,\om+\ol)$ &   $(-0.49\pm0.12, 1.11\pm0.52)$ \\
$(\om,\ol)$ flat &  $\om = 0.263 \pm 0.037$ \\
$(\om,\ol)$ + BAO & $(0.271\pm0.020, 0.751\pm0.082)$\\
\hline
$(\om,w)$+BAO  & $(0.271\pm 0.021, -1.023\pm0.087)$ \\
\hline
\end{tabular}
\caption{Cosmological parameters and statistical errors of 
Hubble diagram fits, with
the BAO prior where applicable.\label{tab:cosmo_fits}}
\end{table}
\end{center}

Using Monte Carlo realizations of our SN sample, we checked that our
estimators of the cosmological parameters are unbiased (at the level
of 0.1 $\sigma$), and that the quoted uncertainties match the observed
scatter.
We also checked the field-to-field variation of the cosmological
analysis. The four $\om$ values (one for each field, assuming $\ok=0$)
are compatible at 37\% confidence level. We also fitted separately the
Ia and Ia* SNLS samples and found results compatible at the 75\%
confidence level.

We derive an intrinsic dispersion, $\sigma_{int}=0.13 \pm0.02$, 
appreciably 
smaller than previously measured~\citep{Riess98b,Perlmutter99,Tonry03, 
Barris04,Riess04}. 
The intrinsic dispersions of nearby only ($0.15\pm0.02$)
and SNLS only ($0.12\pm0.02$) events are statistically consistent although 
SNLS events show a bit less dispersion.

A notable feature of Figure~\ref{fig:hubble} is that the error bars
increase significantly beyond z=0.8, where the $\zme$ photometry is
needed to measure rest-frame $B-V$ colors.
The $\zme$ data is affected by a low signal-to-noise ratio 
because of low quantum efficiency and high sky background.
For $z > 0.8$, $\sigma((B-V)_{restframe})
\simeq 1.6\sigma(\ime-\zme)$, because the lever arm between the central wavelengths of
$\ime$ and $\zme$ is about
1.6 times lower than for $B$ and $V$.  Furthermore, errors in rest-frame color are
scaled by a further factor of $\beta \simeq 1.6$ in the distance
modulus estimate. With a typical measurement uncertainty $\sigma(\zme)
\simeq 0.1$, we have a distance modulus uncertainty $\sigma(\mu) > 0.25$.  
Since the fall 2004 semester, we now acquire about three 
times more $\zme$ data than for the data
in the current paper, and this will improve the accuracy of future cosmological
analyses.

The distance model we use is linear in stretch and color. Excluding
events at $z>0.8$, where the color uncertainty is larger than the 
natural color dispersion, we checked that adding quadratic terms in stretch 
or color to the distance estimator decreases the minimum $\chi^2$ by less 
than 1. We hence conclude that the linear distance estimator accurately
describes our sample.

Since the distance estimator we use depends on the color parameter
$c$, residuals to the Hubble Diagram are statistically correlated to
$c$. The correlation becomes very apparent when the $c$ measurement
uncertainty dominates the distance uncertainty budget, as happens in
our sample when $z>0.8$. We checked that the measurement uncertainties can
account for the observed residual-$c$ correlation at $z>0.8$.  Because
of this correlation, color selected sub-samples mechanically lead to biased
estimations of cosmological parameters.

\section{Comparison of nearby and distant SN properties}
\label{section:comparison}

\subsection{Stretch and color distributions}

\begin{figure}
\begin{center}
\includegraphics[width=\linewidth]{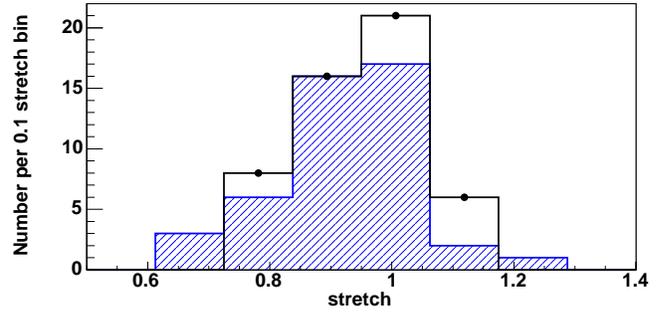}
\caption{The stretch ($s$ parameter) distributions 
of nearby (hashed blue) and distant (thick black with filled symbols) 
SNLS SNe with $z<0.8$. These distributions are
very similar with averages 
of $0.920\pm0.018$ and $0.945\pm0.013$, respectively (1$\sigma$ apart).
\label{fig:stretch_distribution}
}
\end{center}
\end{figure}

\begin{figure}
\begin{center}
\includegraphics[width=\linewidth]{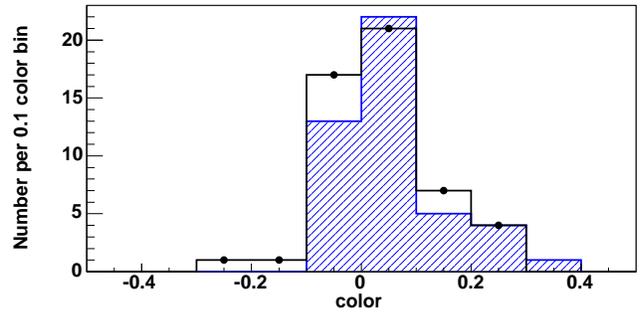}
\caption{The color ($c$ parameter) distributions for 
nearby (hashed blue) and distant (thick black with filled symbols) 
SNe with $z<0.8$. These distributions are very similar, with 
averages of $0.059\pm0.014$ and $0.029\pm0.015$, 
respectively (1.5$\sigma$ apart).
\label{fig:color_distribution}
}
\end{center}
\end{figure}

The distributions of the shape and color parameters -- $s$ and $c$ 
as defined in Section~\ref{section:lightcurve_fit_model} --  
are compared in Figures~\ref{fig:stretch_distribution} and 
\ref{fig:color_distribution} for nearby objects 
and for SNLS supernovae at $z < 0.8$ for which $c$ 
is accurately measured. 
These distributions look very similar, both in central value
and shape. The average values for the two samples differ by about $1\sigma$
in stretch and $1.5\sigma$ in color: we find that
distant supernovae are on average slightly bluer and slower than nearby ones. 
The statistical significance of the differences is low and
the differences can easily be interpreted in terms of selection effects 
rather than evolution.
The evolution of average $s$ and $c$ parameters
with redshift is shown in Section~\ref{section:malmquist_bias};
stretch is not monotonic, and color seems to drift towards the blue
with increasing redshift. We show in 
Section~\ref{section:malmquist_bias} that the bulk of this effect
can be reproduced
by selection effects applied to an unevolving population.

\subsection{Brighter-slower and brighter-bluer relationships}

\begin{figure}
\begin{center}
\includegraphics[width=\linewidth]{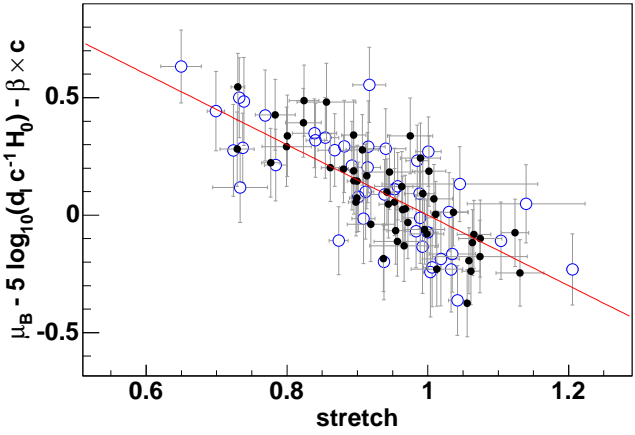}
\caption{Residuals in the Hubble diagram as a function of 
stretch ($s$ parameter), for nearby (blue open
symbols) and distant ($z<0.8$, black filled symbols). This diagram 
computes distance
modulus $\mu_B$ without the stretch term $\alpha (s-1)$, and returns
the well-known brighter-slower relationship with a
consistent behavior for nearby and distant SNe~Ia.
\label{fig:brighter_slower}
}
\end{center}
\end{figure}

\begin{figure}
\begin{center}
\includegraphics[width=\linewidth]{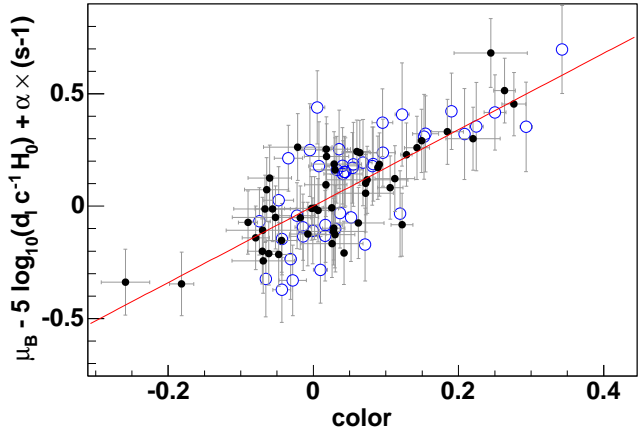}
\caption{Residuals in the Hubble diagram as a function of 
color ($c$ parameter), for nearby (blue open
symbols) and distant ($z<0.8$, black filled symbols). 
This diagram computes distance
modulus $\mu_B$ without the color term $\beta c$, and returns
the brighter-bluer relationship with a consistent 
behavior for nearby and distant SNe~Ia.
Notice that the bluest SNLS objects are compatible with
the bulk behavior.
\label{fig:brighter_bluer}
}
\end{center}
\end{figure}

Figures~\ref{fig:brighter_slower} and
\ref{fig:brighter_bluer} compare the nearby and distant samples
in the stretch-magnitude and color-magnitude planes. 
There is no significant difference between these samples.

In Figure~\ref{fig:color_distribution},
two of the SNLS events (SNLS-04D1ag and SNLS-04D3oe) have a color value, $c$, 
smaller than $-0.1$. These supernovae are both classified as secure Ia.
There are no SNe~Ia in the nearby sample that are this blue.  
Figure~\ref{fig:brighter_bluer} shows that these events lie on the 
derived brighter-bluer relation.
Although they are brighter than average, fitting with or without these
two events changes the cosmological results by less than 0.1 $\sigma$.

\subsection{Compatibility of SN colors}
\label{section:SNcolorcompatibility}

The measurement of distances to high redshift SNLS SNe involves the 
rest-frame $U$
band.  The MegaCam $\rme$ band shifts from  rest-frame $B$ at
z=0.5 to  rest-frame $U$ at z=0.8.  Within this redshift range,
distances are estimated mainly using $\ime$ and $\rme$, the weight of $\zme$
being affected by high photometric noise;
the ($\rme,\ime$) pair roughly changes from rest-frame ($B$,$V$) to rest-frame
($U$,$B$).

Our cosmological conclusions rely on having a consistent
distance estimate when using rest-frame $BV$ and $UB$.
This property is tested in \cite{Guy05}. However, it can be
tested further on the subset of SNLS data having at least three usable 
photometric bands. The test proceeds as follows:
\begin{enumerate}
\item We fit the three bands at once, and store the stretch
and date of maximum $B$ light.
\item We fit the two reddest bands ($BV$ for nearby objects), with
the stretch, and date of maximum being fixed at the previously
obtained values.  From the fitted light-curve model we extract the
expected rest-frame $U$ band magnitude at maximum $B$ light, $U_{BV}$.
\item We fit the two bluest bands, ($UB$ for nearby objects), still with
the stretch and date of maximum fixed. From this fit, we extract the expected
rest-frame $U$ band magnitude at maximum $B$ light. Since it matches the
measurement when the actual $U$ flux is measured, we call it $U_{meas}$.
\end{enumerate}
The test quantity is $\Delta U_3 \equiv U_{BV} - U_{meas}$, i.e. 
the ``predicted'' $U$
(derived from $B$ and $V$) minus the measured $U$ brightness. 
Forcing both quantities to be
measured with the same stretch and $B$ maximum date is not essential,
but narrows the distribution of residuals. A residual of zero means that
the three measured bands agree with the light-curve model for a certain
parameter set, and hence that the
distance estimate will be identical for the two different color fits.

\begin{figure}
\begin{center}
\includegraphics[width=\linewidth]{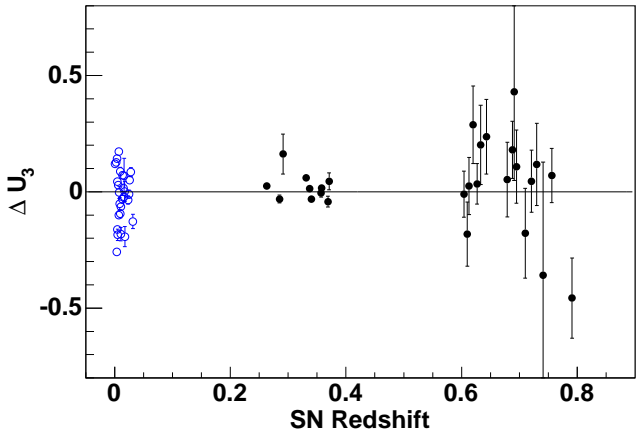}
\caption{$\Delta U_3$, difference between rest-frame $U$ peak 
magnitude ``predicted'' from $B$ and $V$,
and the measured value, as a function of redshift. The error bars
reflect photometric uncertainties. The redshift regions
have been chosen so that the measured bands roughly sample the $UBV$ rest-frame
region. The differences between average values for the three samples
agree within statistical uncertainties, indicating that the relation
between $U$, $B$ and $V$ brightnesses does not change with redshift. Although
the  nearby and intermediate samples have comparable photometric resolution,
the intermediate sample exhibits a far smaller scatter. We attribute this
difference to the practical difficulties in calibrating $U$ band observations.
\label{figure:U-U}}
\end{center}
\end{figure}

There are 10 SNLS ``intermediate'' redshift events at $0.25<z<0.4$, 
where $\gme\rme\ime$ sample the
$UBV$ rest-frame region, and 17 ``distant'' events at $0.55<z<0.8$, 
where $UBV$
shifts to the $\rme\ime\zme$ triplet.  We also have at our disposal
a sample of 28 ``nearby''
objects measured in $UBV$, both from the nearby sample described in
Table~\ref{table:nearbydata}, and also from the light-curve model 
training sample which consists mainly of very nearby 
objects (see~\citealt{Guy05}). 
Figure~\ref{figure:U-U} displays the value
of $\Delta U_3$ as a function of redshift and Table~\ref{table:U-U} 
summarizes the averages and dispersions. A very small scatter
(about 0.033) is found for the intermediate redshift sample. 
The nearby and
distant samples exhibit larger scatters; the nearby sample is probably
affected by the practical difficulties in calibrating $U$ observations,
and our distant sample is affected by the poor S/N in the $\zme$ band.
We conclude from this study that our light curves model accurately describes
the relations between the supernovae colors.
Note that this $\Delta U_3$ indicator is a
promising tool for photometric classification of SNe~Ia, provided its
scatter remains comparable to that found for the intermediate redshift
sample.

\begin{center}
\begin{table}[h]
\begin{tabular}{l|c|c|c|c}
Sample & Bands & Events & r.m.s & Average \\
\hline 
nearby       & UBV                & 28 & 0.122 & 0.0008 $\pm$ 0.023\\
intermediate & $\gme$$\rme$$\ime$ & 10 & 0.033 & 0.009 $\pm$ 0.010 \\
high-z       & $\rme$$\ime$$\zme$ & 17 & 0.156 & 0.039 $\pm$ 0.035 \\
\hline
\end{tabular}
\caption{Statistics of the 3 samples displayed in 
Fig.~\ref{figure:U-U}.\label{table:U-U}}
\end{table}
\end{center}

The same exercise can be done without imposing identical  stretch
and date of maximum light on the two fits. Rather than testing
the light curves model, one then tests
for potential biases in color estimates (leading to biases in distance
estimates). The conclusions are the same
as with fixed parameters: the samples have averages consistent with 0,
and the dispersion of the central sample increases from 0.033 to 0.036.

\section{Systematic uncertainties}
\label{section:systematic_uncertainties}

We present, in this Section, estimates of the systematic uncertainties 
possibly affecting our cosmological parameter measurements.

\subsection{Photometric calibration and filter band-passes}

We simulated a zero-point shift by varying the magnitudes of
the light-curve points, one band at a time. Table \ref{tab:zeropoint_shifts}
gives the resulting shifts in the derived cosmological parameters from
the calibration errors derived in 
Section~\ref{section:calibration_of_field_stars}. 
We assume that
errors in the $\gme\rme\ime\zme$ zero-points are independent, 
and propagate these
4 errors quadratically to obtain the total effect on cosmology.

\begin{center}
\begin{table}[h]
\begin{tabular}{c|cccc}
Band & zero-point shift & $\delta \om$ (flat) & $\delta \ot$ & $\delta w$ (fixed$\om$) \\
\hline 
$\gme$ & 0.01 & 0.000  & -0.02     &  0.00 \\
$\rme$ & 0.01 & 0.009  &  0.03     &  0.02 \\
$\ime$ & 0.01 & -0.014 &  0.17     &  -0.04 \\
$\zme$ & 0.03 & 0.018  &  -0.48    & -0.03 \\
\hline
sum &  -    & 0.024  &  0.51     & 0.05 \\
\hline
\end{tabular}
\caption{Influence of a photometric calibration error on the 
cosmological parameters.\label{tab:zeropoint_shifts}}
\end{table}
\end{center}

We rely on the spectrum of one object, Vega ($\alpha$ Lyrae), to transform magnitudes into
fluxes; the broadband flux errors for Vega are about 1\% (\citet{Hayes85} and
Section~\ref{section:converting_magnitudes_to_fluxes}). To take
into account the Vega flux and broadband color uncertainties, 
we simulated a flux error linear in wavelength that would
offset the Vega ($B-R$) color by 0.01. The impact on $\om$ is $\pm$0.012.

Uncertainties in the filter bandpasses affect the determination of
supernovae brightnesses; the first-order effect is from errors in the central
wavelengths. In the color-color relations (Landolt/MegaCam and
SDSS/MegaCam -- Section~\ref{section:instrumental_filters}), 
we were able to detect shifts of 10~\AA\ (corresponding
roughly to a change of 0.01 in the color term). The effect of this
shift is in fact very small: only the $\rme$ filter has a sizable
impact of $\pm$0.007 on $\om$.

\subsection{Light-curve fitting, (U-B) color and k-corrections}

If the light-curve model fails to properly describe the true 
light-curve shape, the result would be a bias in the light-curve parameters, 
and possibly
in the cosmological parameters if the bias depends on redshift.  We have 
already discussed two possible causes of such a bias: the influence of the
first measurement date (Section~\ref{section:snIa-sample}), 
and the choice of rest-frame
bands used to measure brightness and color
(Section~\ref{section:SNcolorcompatibility}). Both have very small 
effects. However,
given only 10 intermediate redshift SNLS events, each with an uncertainty 
of $0.033$, the precision with which we
can define the average ($U-B$) color at given ($B-V$) is limited to about
0.01~mag by our sample size.

Uncertainties in the k-corrections
(due to SNe~Ia spectral variability at fixed color)
contribute directly to the observed scatter. The redshift range of
the intermediate redshift sample of 
Section~\ref{section:SNcolorcompatibility} 
corresponds to a 
rest-frame wavelength span of about 400 \AA, in a region where 
SNe~Ia spectra are highly structured.
Since we observe compatible intrinsic dispersions for nearby and SNLS events 
(indeed, slightly lower for SNLS), we find no evidence that
k-correction uncertainties add significantly to the intrinsic dispersion.

Nevertheless, since the measured scatter of the intermediate redshift sample 
appears surprisingly small and, since the sample size is small,
we used a more conservative value of 0.02 for the light-curve model
error, to account for both the errors in the colors and from k-corrections.
A shift of the $U$-band light-curve model of 0.02 mag results in a
change in $\om$ of 0.018. This is to be added to the statistical
uncertainty.

\subsection{$U$-band variability and evolution of SNe~Ia}

Concerns have been expressed regarding the use 
of rest-frame $U$-band fluxes to measure luminosity
distances (e.g. \citealt{Jha02} and \citealt{Nugent02}), 
motivated by the apparent large variability of the $U$-band
luminosity of SNe~Ia. Such variability seems also to be present
at intermediate redshifts although there seems to be little obvious 
evolution to $z=0.5$ of the overall UV SED 
(\citeauthor{Ellis05}, in prep.).
Note that \cite{Guy05} have succeeded in
constructing a distance estimator using $U$ and $B$-band data 
which shows a dispersion of only 0.16 mag around the Hubble line,
comparable to that found for distances derived using $B$- and $V$-band data.
Note also that the quantity $\Delta U_3$ appears to be independent of
redshift, implying that if the average luminosity of SNe~Ia evolves with 
redshift, this evolution must preserve the $UBV$ rest-frame color relations.  
\cite{Lentz2000} predict a strong
dependence of the UV flux from the progenitor metallicity 
(at fixed $B-V$ color), which
should have been visible if metallicity evolution were indeed present.

\subsection{Malmquist bias\label{section:malmquist_bias}}

The Malmquist bias may affect the cosmological conclusions by altering
the average brightness of measured SNe in a redshift dependent way.
The mechanism is however not exactly straightforward since the reconstructed
distance depends on stretch and color, and not only on the brightness.
We have conducted simulations, both of nearby SN searches and of the SNLS
survey, to investigate the effects on the derivation
of cosmological parameters.

We simulated light-curves of nearby SNe~Ia ($0.02<z<0.1$) with
random explosion date, stretch and color, using the observed brighter-slower
and brighter-bluer correlations. We then simulated a brightness cut at
a fixed date.  Although the number of ``detected'' events and their
average redshift strongly depends on the brightness cut, the average
distance bias of the survivors is found to change by less than 10\%,
when varying both the value and the sharpness of the brightness cut. 
The bias is
also essentially independent of the discovery phase, although the peak
brightness is not.  We find a distance
modulus bias of 0.027 (similar in $B$, $V$ and $R$), sensitive at the 10\% level to the
unknown details of nearby searches. Note that the redshift dependence of
the distance bias of the nearby sample has no impact on the
cosmological measurements: only the average bias matters.

The crude simulation we conducted  applies only to flux limited
searches, which applies to about half of the sample.
We compute an average bias value for our nearby sample as the simulation result
(0.027 mag) times the fraction of events to which it applies. 
Assuming that both
factors suffer from an uncertainty of 50\%, we find an average nearby
sample bias value of $0.017 \pm 0.012$ mag. A global increase of all
nearby distances by 0.017($\pm 0.012$) mag increases $\om$ (flat universe) by
0.019 ($\pm 0.013$).

\begin{figure}
\begin{center}
\begin{minipage}[c]{0.49\linewidth}
\includegraphics[width=\linewidth]{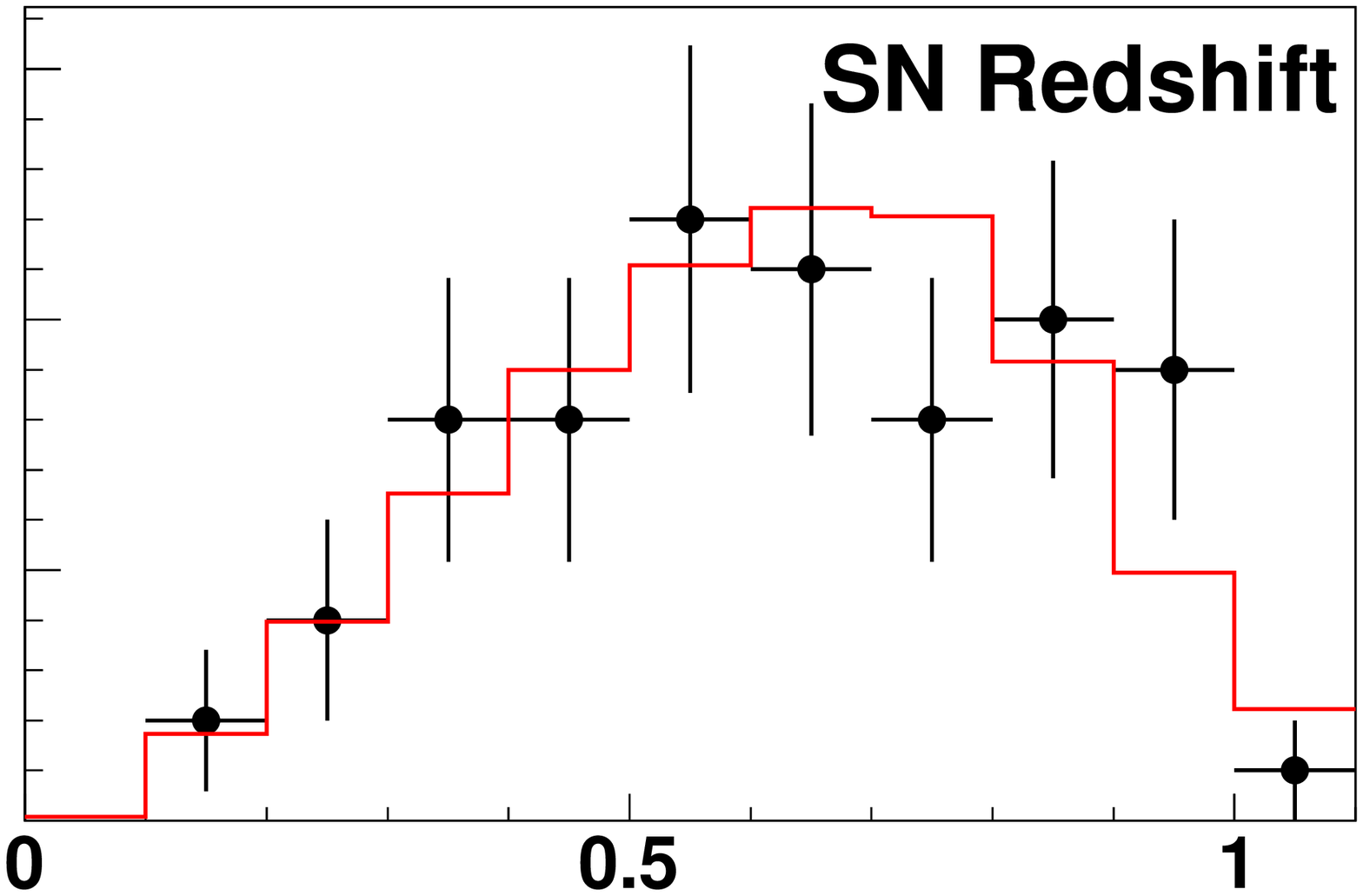}
\end{minipage}\hfill
\begin{minipage}[c]{0.49\linewidth}
\includegraphics[width=\linewidth]{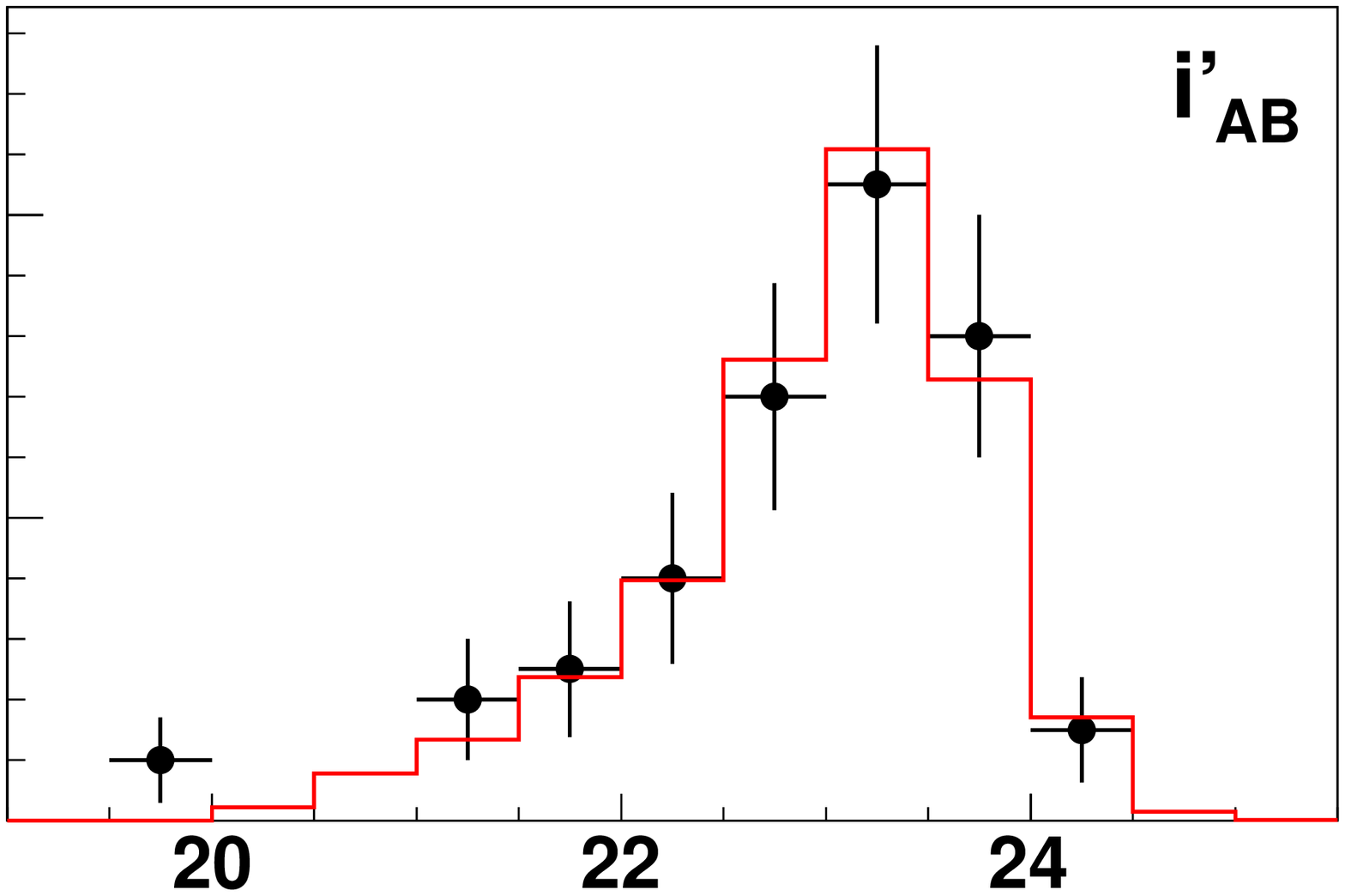}
\end{minipage}
\begin{minipage}[c]{0.49\linewidth}
\includegraphics[width=\linewidth]{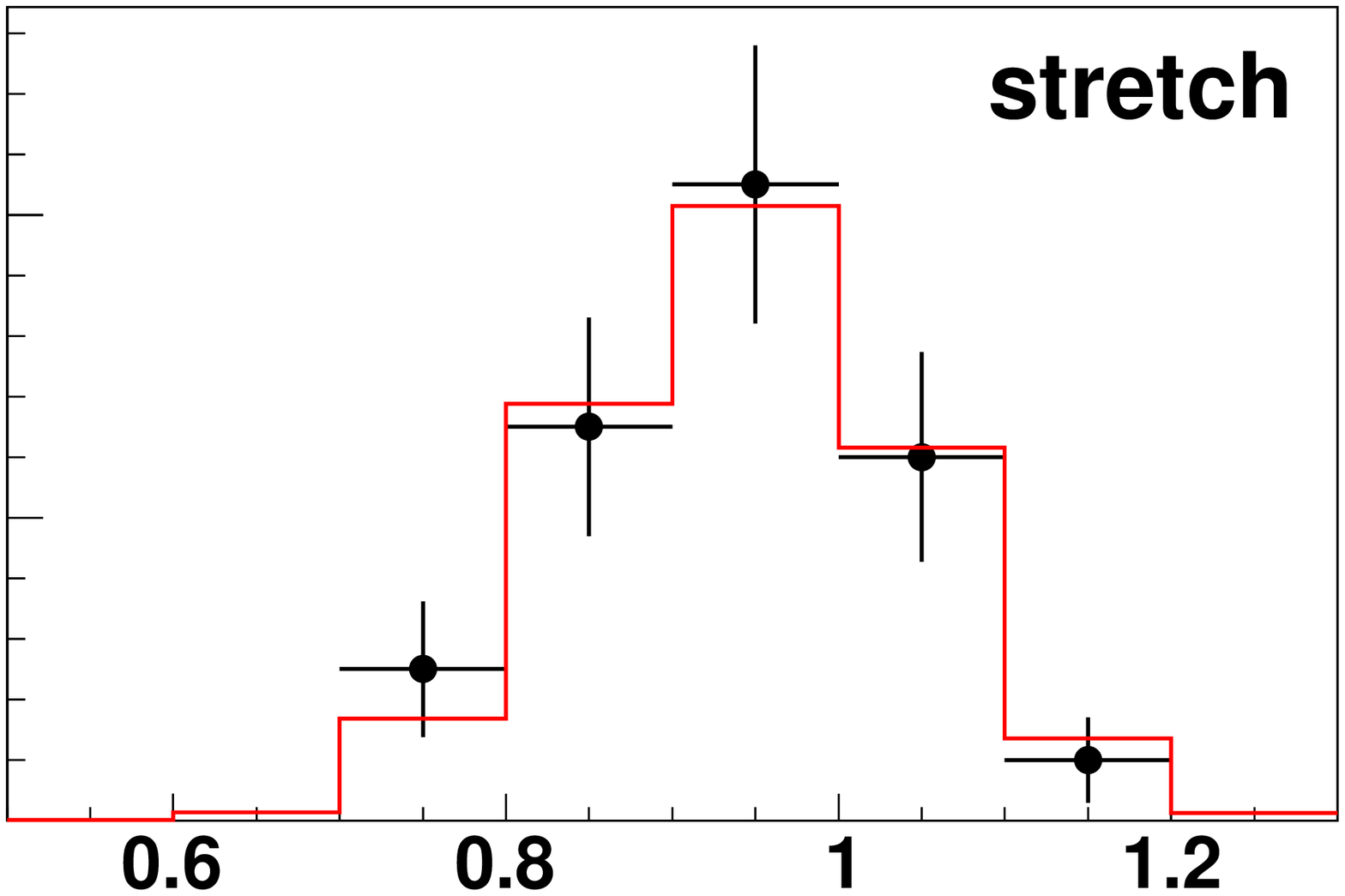}
\end{minipage}\hfill
\begin{minipage}[c]{0.49\linewidth}
\includegraphics[width=\linewidth]{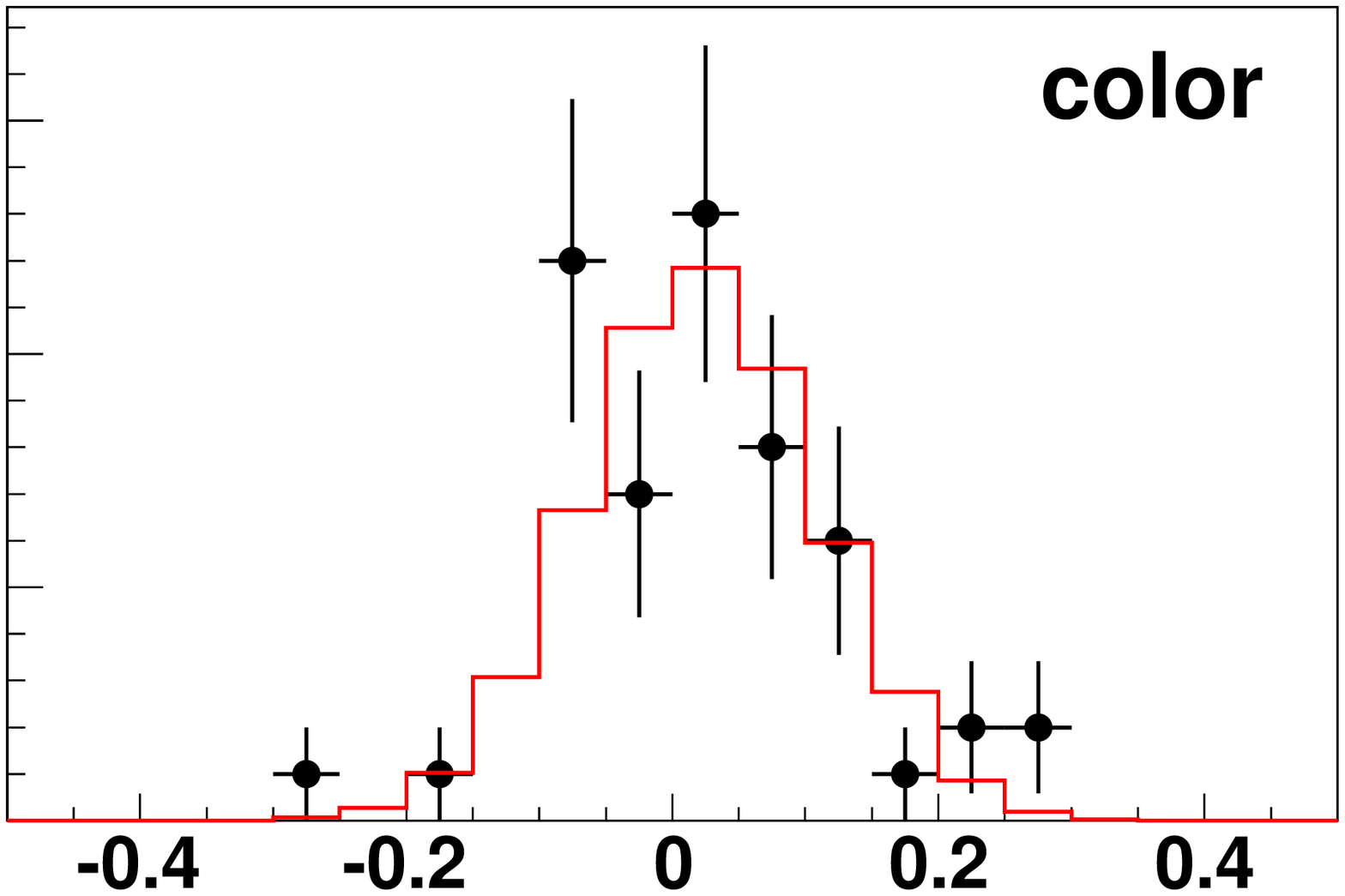}
\end{minipage}
\caption{Distributions of redshifts, peak $\ime$ magnitudes (AB), 
stretch factors and colors of SNLS supernovae (black dots) together 
with the distributions  obtained with simulated SNe (red histograms). 
\label{figure:malm_snls_1}
}
\end{center}
\end{figure}

\begin{figure}
\begin{center}
\includegraphics[width=\linewidth]{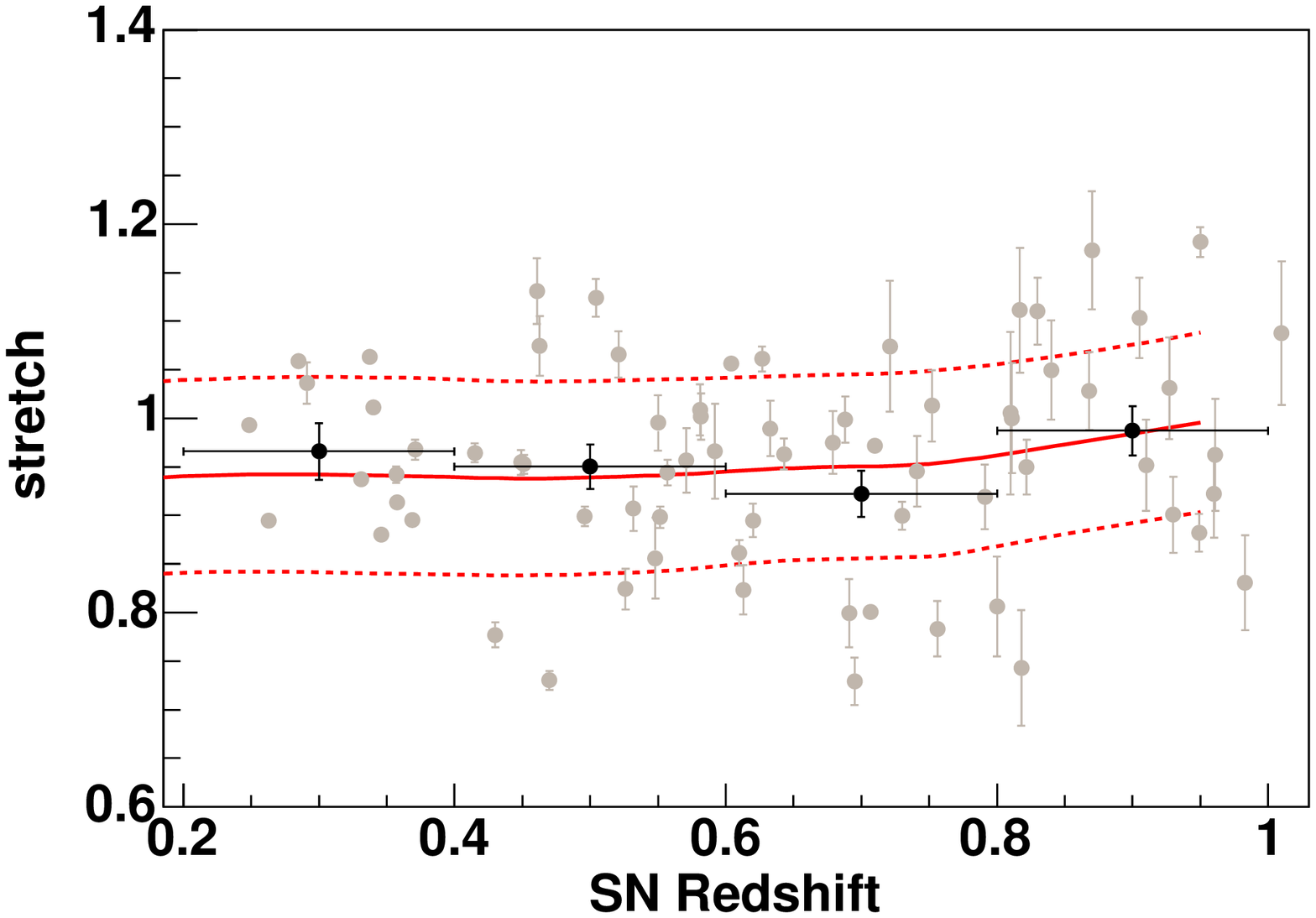}
\includegraphics[width=\linewidth]{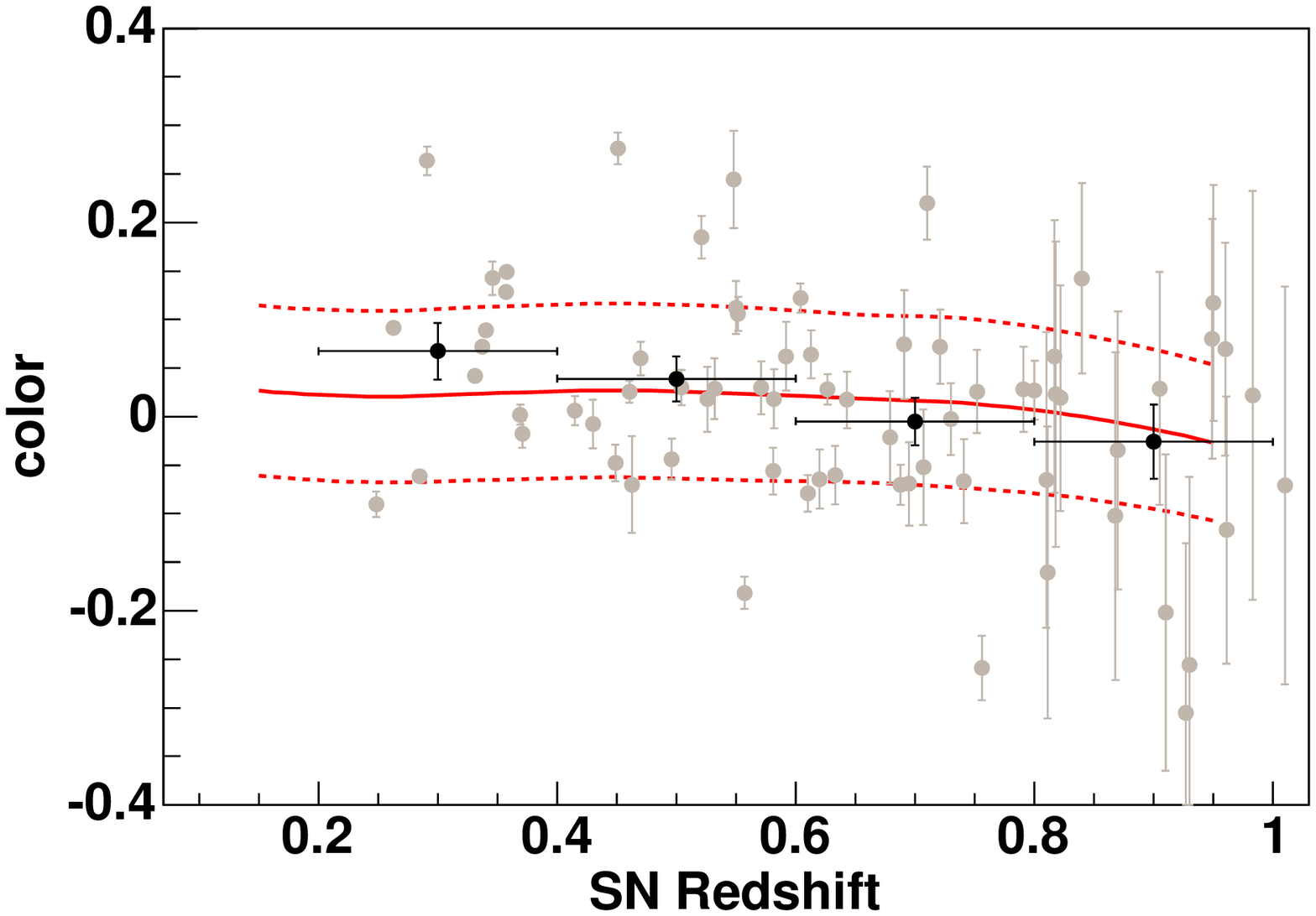}
\includegraphics[width=\linewidth]{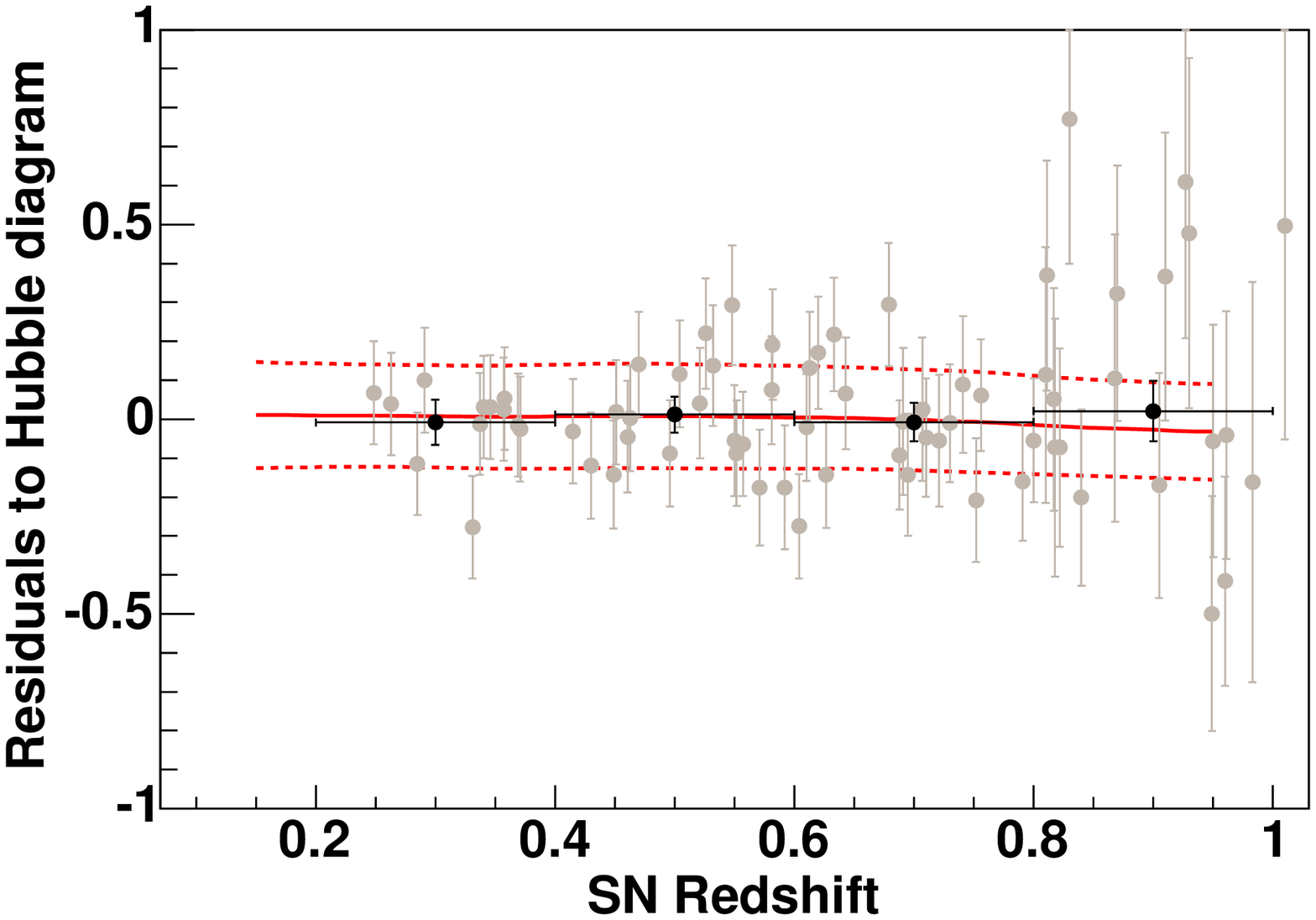}
\caption{
\label{figure:malm_snls_2}
Stretch, color and Hubble diagram residuals as a function of redshift 
for SNLS supernovae (gray dots). The black points correspond to average 
values in redshift bins. The red solid (dashed) lines represent the average 
(one standard deviation) values obtained with SNe simulations as
described in 
Section~\ref{section:malmquist_bias}. 
At large redshifts, 
since only bright SNe are identified, the average stretch factor is larger 
and the average color bluer. The average distance modulus is less affected by 
the selection (see text for details).
}
\end{center}
\end{figure}

For the distant SNLS sample, which is flux limited,
we simulated supernovae at a rate per co-moving volume 
independent of redshift, accounted for the
brighter-slower and brighter-bluer correlations, and adjusted the
position and smoothness of the limiting magnitude cut in order to reproduce the
redshift and peak magnitude distributions. In contrast with nearby SN
simulations, here we have many observed distributions for a single search,
and the key parameters that enter the simulation are highly constrained.
The best match to SNLS data is shown in 
Figure~\ref{figure:malm_snls_1},  
and Figure~\ref{figure:malm_snls_2} 
shows the expected biases as a function of redshift in the shape 
and color parameters, and for our distance estimator. 
The distance modulus bias is about 0.02 mag at
$z=0.8$,  increasing to 0.05 at $z=1$. Correcting for the computed
bias decreases $\om$ (flat Universe) by 0.02. We assumed that the
uncertainty in this bias correction is 50\% of its value.

To summarize: we find that the differential bias between nearby and
distant samples almost exactly cancels, and estimate an overall
uncertainty of 0.016 in $\om$ (flat Universe). Since applying the
Malmquist bias corrections changes the cosmological results by less than 0.1
$\sigma$, the corrections have not been applied. However, in the future,
when the SNLS sample size increases, modeling and applying the
Malmquist bias correction will assume a greater importance. The same
applies to the
nearby sample, where having a more controlled and homogeneous sample, 
discovered by
a single search (e.g. SN Factory,~\citealt{SNFAldering}) 
will be essential to reduce the associated systematic uncertainty.

\subsection{SNe~Ib/c interlopers}

All supernovae used here were spectroscopically identified as SN~Ia, but we 
have labeled the least secure identifications as SN Ia*
(Section~\ref{section:spectro_follow_up}, \citealt{Howell05}).
These 15 events are probable SN~Ia but for this class a small amount 
of contamination
by SNe~Ib or SNe~Ic (SNe~Ib/c) is possible. We have checked that 
cosmological fits done with or without these events
lead to the same cosmological conclusions
(Section~\ref{section:cosmo-fit}).  

We also looked at estimating the SN Ib/c contamination in our sample. 
SNe~Ib/c have an intrinsic
luminosity distribution which is wider than SNe~Ia
(cf. dispersion 0.45 mag for SNe~Ia, vs. 1.2 mag for SNe~Ib/c;
\citealt{Homeier05,Richardson02}).  After correcting for the SNe~Ia
brighter-slower and brighter-bluer correlations, a conservative estimate is
that the SNe~Ib/c scatter around the SNe~Ia Hubble line with a dispersion
3 to 4 times larger
than for SNe~Ia. The first clue of SN~Ib/c contamination would be the
presence of objects with large residuals around the Hubble line; these
contaminants should on average be fainter than SNe~Ia at the same redshift.
We have rejected two objects from the Hubble diagram
(Section~\ref{section:cosmo-fit}). Even if we consider
both of these events to be SN~Ib/c events,
and assume that the dispersion of the
SN~Ia distribution about the Hubble line is 4 times smaller 
than for SNe~Ib/c, we expect on average only 0.5 Ib/c interloper within the 
fitted sample. 

For these reasons, we estimate the potential bias arising from the 
presence of non Ia events in our sample to be negligible. 

\subsection{Gravitational lensing and grey dust}
Gravitational lensing by mass inhomogeneities may affect the apparent
brightness of our supernovae. With respect to a uniform matter density,
most of the events experience a tiny de-amplification, and a small
minority are amplified (see e.g. \citealt{HolzWald98}).

Whereas the average flux is conserved in the case of weak lensing,
part of the SN light is lost when strong lensing produces multiple
images among which some escape detection.  Multiple images of distant
radio sources have been systematically searched by the CLASS
survey~\citep{ClassSurvey} and have proved to be rare: the occurrence
of multiple images separated by more than $0.3\arcsec$ and with flux
ratio below 10:1 was found to be of $1$ out of $690$
with 1.44 secondary image on average, 
with inefficiencies due to the
separation and flux ratio cuts of 13\% and 37\%
respectively \citep{Browne03}.
Multiple images with a smaller separation are not
resolved in the SNLS, and their time delay is much smaller than the
typical duration of a SN light curve\footnote{Delays are of order of a day
for a source at $z=1$ and a point-like lens at $z=0.5$ for a typical
angular separation of 0.2\arcsec~\citep{Bergstrom00}} so that no flux is
lost for such events. Hence CLASS results provide us with an upper 
limit for the number of (resolved) strong lensing cases in the SNLS
supernova sample, given the fact that CLASS sources are globally more
distant~\citep[see][]{Chae03}. Assuming (pessimistically) that for each
strongly lensed SN, we see only one image, the flux bias is smaller than
$0.3\%$ at $z=1$.

Gravitational lensing also broadens asymmetrically the brightness
distribution of SNe at large redshifts~\citep{Bergstrom00}.  As a
consequence, a cosmological fit using SNe magnitudes (instead of
fluxes) is biased. \citet{HolzLinder04} found a dispersion of $0.088
\times z$ (note that \citealt{Bergstrom00} find a value of $\sim$0.04 at $z=1$
for smooth halo profiles in flat $\Lambda$CDM), which translates into a bias
of the average magnitude of $\sim\,0.004 \times z$.  The broadening of
the brightness distribution also affects the cosmological parameters
uncertainties. In the cosmological fit, we have derived a constant
``intrinsic'' dispersion which includes the average dispersion due to lensing.
Neglecting its redshift dependence has no significant impact on the
accuracy of the errors derived for the cosmological parameters.

In summary, the total effect of lensing on cosmological parameters is
very small. We find that $\om$ for a flat $\Lambda$CDM cosmology 
and the equation of state for a flat
universe with BAO constraints, are shifted
by at most $-0.005$ and $-0.01$ respectively. We
therefore did not apply any correction to our results.

The possibility that SNe~Ia could be dimmed by intergalactic grey dust 
(i.e. with weak extinction variation over the optical wavelengths)
has been suggested by \citet{Aguirre99a,Aguirre99b} as an astrophysical
alternative to the dark energy hypothesis. Some simple dust scenarios
without a cosmological constant could be excluded by \cite{Riess04}
using SNe~Ia data.  Studying the colors of a large sample of quasars,
\cite{qsocol2} were able to set limits on the light absorption length
as a function of $R_V$, but these limits can only be translated into an
upper bound of supernovae dimming. Conservatively assuming $R_V = 12$, 
using the SNOC program~\citep{SNOC}, we computed an upper
limit in the dimming of supernovae which translates into 
a shift of $-0.025$ in $\om$ for a $\Lambda$CDM cosmology, and a shift
of $-0.048$ in $w$ for a flat cosmology with constant equation of state when 
combined with SDSS BAO results.
Note that these are upper limits and that a scenario without any intergalactic 
dust cannot be excluded. We therefore did not apply any correction to our 
results.

\section{Summary and perspectives}

Table \ref{tab:uncertainties} summarizes the uncertainties affecting
our cosmological parameter measurements. The table includes the impact of
uncertainties in several parameter directions: the $\om$ direction for a flat
$(\om,\ol)$ (i.e. $w=-1$) cosmology, the $\ot$ direction for a general
$(\om,\ol)$ cosmology, and the $w$ direction at fixed $\om$ for a $(\om,w)$
cosmology. We also report here the observed shifts when the BAO prior 
is applied to a flat $(\om,w)$ cosmology.
\begin{center}
\begin{table}[h]
\begin{tabular}{c|ccc|cc}
Source & $\sigma(\om$)  & $\sigma(\ot)$ & $\sigma(w)$ & $\sigma(\om)$  &  $\sigma(w)$\\
       & (flat)        & & & \multicolumn{2}{c}{(with BAO)}\\
\hline 
\hline
Zero-points               &  0.024   &   0.51    &  0.05    & 0.004  & 0.040    \\
Vega spectrum             &  0.012   &   0.02    &  0.03    & 0.003  & 0.024    \\
Filter bandpasses         &  0.007   &   0.01    &  0.02    & 0.002  & 0.013    \\
Malmquist bias            &  0.016   &   0.22    &  0.03    & 0.004  & 0.025    \\
\hline
Sum (sys)                 &  0.032   &   0.55    &  0.07    & 0.007  & 0.054    \\
\hline 
\hline
Meas. errors        &  0.037   &   0.52    &  0.09    & 0.020  & 0.087    \\
U-B color(stat)           &  0.020   &   0.10    &  0.05    & 0.003  & 0.021    \\
\hline
Sum (stat)                &  0.042   &   0.53    &  0.10    & 0.021  & 0.090    \\
\hline
\hline 
\end{tabular}
\caption{
Summary of uncertainties in the derived cosmological parameters. The dominant
systematic uncertainty arises from the photometric calibration, itself
dominated by the $\ime$ and $\zme$ band contributions.
\label{tab:uncertainties}
}
\end{table}
\end{center}
Note that measurement and isolation of systematic errors is a 
major goal of the SNLS. Some of these uncertainties will decrease as more
data is acquired and future papers will examine a wider range of issues,
using our growing dataset.

Combining Tables~\ref{tab:cosmo_fits} 
and \ref{tab:uncertainties}, we obtain the following results:

$$\om = 0.263 \pm 0.042\;(stat) \pm 0.032\;(syst)$$ 

\noindent
for a flat $\Lambda$ cosmology, and
\begin{center} 
\begin{tabular}{rcl}
$\om$ & = & $ 0.271 \pm 0.021\;(stat) \pm 0.007\;(syst) $ \\
$w  $ & = & $ -1.023 \pm 0.090\;(stat) \pm 0.054\;(syst) $ \\
$w  $ & $<$ & $ -0.85 \ \ (95\%\ \ CL)$
\end{tabular}
\end{center}
for a 
flat cosmology with constant equation of state, when our constraints
are combined with the BAO SDSS results. Assuming $w>-1$ brings
our upper limit to $-0.83$ (at 95\%~CL). Supernovae alone give a
marginal constraint: $w<-0.5$ at 95\%~CL.

These results agree well with previous works, both from
SNe~Ia, and also from other sources. For example~\cite{Seljak05} finds
very similar results combining CMB, LSS and Ly$\alpha$ constraints.
The dominant systematic errors arise from the nearby
sample and from the photometric calibration of the $\zme$ band; both will be
improved in the future. The multi-band light-curves allow us to study
color relations as a function of redshift; these data are expected to be
sensitive indicators of evolution. We observed a surprisingly narrow
correlation between (U-B) and (B-V) (using the $\Delta U_3$ indicator),
indicating that the dispersion in $U$-band properties is well correlated with
measurements in redder bands.

From the first year of SNLS data, we placed 71 distant events on the
Hubble Diagram, with 10 more from the same period to be added later. 
(Our full first year statistics would have been around 100 SNe~Ia with
spectroscopic confirmation had we not lost Feb 2004 to an instrument
failure.) Our time sampling, filter coverage, and image quality have
now significantly improved since early 2004,
and we now regularly acquire 2--3 times as much data in $\zme$.
A precise photometric calibration is essential, and we are
now working with the CFHTLS community in refining
the photometric calibration of the MegaCam instrument.  We have
embarked on the process of calibrating tertiary standards in our
fields, from Sloan secondary and primary standards. This will allow us
to cross-check the Vega/Landolt zero-points, and 
more accurately calibrate $\zme$ band observations.\\

After only two years of operation, the SNLS has already demonstrated
its advantages over all previous ground-based supernova surveys.  The
"rolling search" technique is robust to weather and instrument-related
problems, and the technical characteristics of the survey are now well
understood.  The average rate of spectroscopically-confirmed SNe~Ia is
currently about 10 per lunation and continues to increase.  Up until
July 2005, the SNLS sample includes more than 200 spectroscopically identified
SNe~Ia, most with excellent photometric temporal and filter coverage.
An extrapolation of the current rate to the end of the survey
indicates that we should reach our goal of building a Hubble diagram
with about 700 spectroscopically identified well-measured SN~Ia
events.  The SNLS already has the largest-ever sample of high-$z$ SNe
discovered by a single telescope, and will eventually produce a
homogeneous, high-quality sample that is an order of magnitude larger
still.

High statistical accuracy benefits the control of systematics. With our
unmatched SN statistics, by year 5
we will be able to populate each $\sim0.1$ redshift bin 
with $\sim 100$ SNe\,Ia, 
thus filling the brightness, decline-rate, and
color 3-dimensional parameter space. This will enable us to detect possible
drifts in ``SNe~Ia demographics'', and control Malmquist bias. 
Moreover, the rolling-search observing mode produces many
events at low to intermediate redshift with superb photometric
accuracy, because integration times are 
tailored for the faintest objects. These relatively bright events
permit demanding internal consistency tests, 
and may lead to improvements in distance estimation.

\begin{acknowledgements}
The authors wish to recognize and acknowledge the very significant
cultural role and reverence that the summit of Mauna Kea has always
had within the indigenous Hawaiian community.  We are most fortunate
to have the opportunity to conduct observations from this mountain.
We gratefully acknowledge the assistance of the CFHT Queued Service
Observing Team, led by P. Martin (CFHT). We heavily rely on the
dedication of the CFHT staff and particularly J.-C.~Cuillandre for
continuous improvement of the instrument performance. The real-time
pipelines for supernovae detection run on computers integrated in the
CFHT computing system, and are very efficiently installed, maintained
and monitored by K.~Withington (CFHT). We also heavily rely on the
real-time Elixir pipeline which is operated and monitored by
J-C.~Cuillandre, E. Magnier and K.~Withington. We are grateful to
L.~Simard (CADC) for setting up the image delivery system and his kind
and efficient responses to our suggestions for improvements. The
French collaboration members carry out the data reductions using the
CCIN2P3.  Canadian collaboration members acknowledge support from
NSERC and CIAR; French collaboration members from CNRS/IN2P3,
CNRS/INSU, PNC and CEA. This work was supported in part by the
Director, Office of Science, Office of High Energy and Nuclear
Physics, of the US Department of Energy.  The France-Berkeley Fund
provided additional collaboration support. CENTRA members were
supported by Funda\c{c}\~ao para a Ci\^encia e Tecnologia (FCT),
Portugal under POCTI/FNU/43423.  S. Fabbro and \,C. Gon\c{c}alves
acknowledge support from FCT under grants no SFRH/BPD/14682/2003 and
SFRH/BPD/11641/2002 respectively.

\end{acknowledgements}


\bibliographystyle{aa}

\input{cosmo.bbl}

\pagebreak
\include{wholesnlsIa}

\include{nearbydata}

\include{snlsdata}

\appendix

\end{document}

%% file: wholesnlsIa.tex
 
\begin{deluxetable}{lccrlllll}
\tablewidth{0pt}
\tablecaption{Transients from the first year sample identified as SNIa or SNIa*\label{table:wholesnlsIa}}
\tablehead{
\colhead{Name} &
\colhead{RA(2000)} &
\colhead{Dec(2000)} &
\colhead{Obs. date\tablenotemark{a}} &
\colhead{Tel./Inst.\tablenotemark{b}} &
\colhead{Spectral Type\tablenotemark{c}} &
\colhead{z\tablenotemark{d}} &
\colhead{z source} &
}
\startdata
SNLS-03D1au & 02:24:10.392 & -04:02:14.93 &   2907 & Keck/LRIS & SNIa &  0.504 & gal\\
SNLS-03D1aw & 02:24:14.786 & -04:31:01.61 &   2907 & Keck/LRIS & SNIa &  0.582 & gal\\
SNLS-03D1ax & 02:24:23.338 & -04:43:14.28 &   2913 & Gem/GMOS & SNIa &  0.496 & gal\\
SNLS-03D1bf & 02:24:02.375 & -04:55:57.27 &   2909 & VLT/FORS1 & SNIa*  &  0.703 & gal\\
SNLS-03D1bk & 02:26:27.412 & -04:32:11.95 &   2912 & Gem/GMOS & SNIa &  0.865 & gal\\
SNLS-03D1bp & 02:26:37.714 & -04:50:19.55 &   2910 & VLT/FORS1 & SNIa* &  0.346 & gal\\
SNLS-03D1cm & 02:24:55.294 & -04:23:03.61 &   2940 & Gem/GMOS & SNIa &  0.87 & SN\\
SNLS-03D1co & 02:26:16.252 & -04:56:05.65 &   2966 & Keck/LRIS & SNIa &  0.679 & gal\\
SNLS-03D1dj & 02:26:19.087 & -04:07:08.89 &   2964 & Keck/LRIS & SNIa &  0.39 & SN\\
SNLS-03D1dt & 02:26:31.200 & -04:03:08.51 &   2974 & VLT/FORS1 & SNIa  &  0.612 & gal\\
SNLS-03D1ew & 02:24:14.079 & -04:39:56.93 &   2995 & Gem/GMOS & SNIa &  0.868 & gal\\
SNLS-03D1fb & 02:27:12.875 & -04:07:16.44 &   2641 & VLT/FORS1 & SNIa  &  0.498 & gal\\
SNLS-03D1fc & 02:25:43.625 & -04:08:38.93 &   2641 & VLT/FORS1 & SNIa  &  0.331 & gal\\
SNLS-03D1fl & 02:25:58.329 & -04:07:44.17 &   2641 & VLT/FORS1 & SNIa  &  0.688 & gal\\
SNLS-03D1fq & 02:26:55.683 & -04:18:08.10 &   2998 & Gem/GMOS & SNIa &  0.80 & SN\\
SNLS-03D1gt & 02:24:56.027 & -04:07:37.11 &   2641 & VLT/FORS1 & SNIa  &  0.55 & SN\\
SNLS-03D3af & 14:21:14.955 & +52:32:15.68 &   2737 & Keck/LRIS & SNIa &  0.532 & gal\\
SNLS-03D3aw & 14:20:53.534 & +52:36:21.04 &   2767 & Keck/LRIS & SNIa &  0.449 & gal\\
SNLS-03D3ay & 14:17:58.448 & +52:28:57.63 &   2766 & Keck/LRIS & SNIa &  0.371 & gal\\
SNLS-03D3ba & 14:16:33.465 & +52:20:32.02 &   2766 & Keck/LRIS & SNIa &  0.291 & gal\\
SNLS-03D3bb & 14:16:18.920 & +52:14:53.66 &   2766 & Keck/LRIS & SNIa &  0.244 & gal\\
SNLS-03D3bh & 14:21:35.894 & +52:31:37.86 &   2766 & Keck/LRIS & SNIa &  0.249 & gal\\
SNLS-03D3bl & 14:19:55.844 & +53:05:50.91 &   2792 & Keck/LRIS & SNIa &  0.355 & gal\\
SNLS-03D3cc & 14:19:45.192 & +52:32:25.76 &   2793 & Keck/LRIS & SNIa &  0.463 & gal\\
SNLS-03D3cd & 14:18:39.963 & +52:36:44.22 &   2792 & Keck/LRIS & SNIa &  0.461 & gal\\
SNLS-03D4ag & 22:14:45.806 & -17:44:22.95 &   2824 & Keck/LRIS & SNIa &  0.285 & gal\\
SNLS-03D4at & 22:14:24.023 & -17:46:36.03 &   2826 & VLT/FORS1 & SNIa  &  0.633 & gal\\
SNLS-03D4au & 22:16:09.917 & -18:04:39.19 &   2826 & VLT/FORS1 & SNIa*  &  0.468 & gal\\
SNLS-03D4bc & 22:15:28.143 & -17:49:48.66 &   2826 & VLT/FORS1 & SNIa  &  0.572 & gal\\
SNLS-03D4cj & 22:16:06.658 & -17:42:16.83 &   2879 & Keck/LRIS & SNIa &  0.27 & SN\\
SNLS-03D4cn & 22:16:34.600 & -17:16:13.55 &   2879 & Gem/GMOS & SNIa &  0.818 & gal\\
SNLS-03D4cx & 22:14:33.754 & -17:35:15.35 &   2885 & VLT/FORS1 & SNIa  &  0.95 & SN\\
SNLS-03D4cy & 22:13:40.441 & -17:40:54.12 &   2909 & Gem/GMOS & SNIa &  0.927 & gal\\
SNLS-03D4cz & 22:16:41.845 & -17:55:34.40 &   2910 & Gem/GMOS & SNIa &  0.695 & gal\\
SNLS-03D4dh & 22:17:31.040 & -17:37:46.98 &   2906 & Keck/LRIS & SNIa &  0.627 & gal\\
SNLS-03D4di & 22:14:10.249 & -17:30:24.18 &   2885 & VLT/FORS1 & SNIa  &  0.905 & gal\\
SNLS-03D4dy & 22:14:50.513 & -17:57:23.24 &   2912 & VLT/FORS1 & SNIa  &  0.60 & SN\\
SNLS-03D4fd & 22:16:14.462 & -17:23:44.33 &   2937 & Gem/GMOS & SNIa &  0.791 & gal\\
SNLS-03D4gf & 22:14:22.907 & -17:44:02.49 &   2641 & VLT/FORS1 & SNIa*  &  0.58 & SN\\
SNLS-03D4gg & 22:16:40.185 & -18:09:51.82 &   2641 & VLT/FORS1 & SNIa  &  0.592 & gal\\
SNLS-03D4gl & 22:14:44.183 & -17:31:44.36 &   2966 & Keck/LRIS & SNIa &  0.571 & gal\\
SNLS-04D1ag & 02:24:41.125 & -04:17:19.66 &   2641 & VLT/FORS1 & SNIa  &  0.557 & gal\\
SNLS-04D1aj & 02:25:53.982 & -04:59:40.50 &   2641 & VLT/FORS1 & SNIa*  &  0.72 & SN\\
SNLS-04D1ak & 02:27:33.399 & -04:19:38.73 &   2641 & VLT/FORS1 & SNIa*  &  0.526 & gal\\
SNLS-04D2ac & 10:00:18.923 & +02:41:21.63 &   2641 & VLT/FORS1 & SNIa*  &  0.348 & gal\\
SNLS-04D2ae & 10:01:52.361 & +02:13:21.27 &   3026 & Gem/GMOS & SNIa &  0.843 & gal\\
SNLS-04D2al & 10:01:52.482 & +02:09:51.25 &   2641 & VLT/FORS1 & SNIa  &  0.84 & SN\\
SNLS-04D2an & 10:00:52.332 & +02:02:28.73 &   2641 & VLT/FORS1 & SNIa  &  0.62 & SN\\
SNLS-04D2bt & 09:59:32.725 & +02:14:53.07 &   3085 & VLT/FORS1 & SNIa  &  0.220 & gal\\
SNLS-04D2ca & 10:01:20.514 & +02:20:21.76 &   2641 & VLT/FORS1 & SNIa  &  0.83 & SN\\
SNLS-04D2cf & 10:01:56.110 & +01:52:46.40 &   3086 & VLT/FORS1 & SNIa  &  0.369 & gal\\
SNLS-04D2cw & 10:01:22.787 & +02:11:55.31 &   3085 & VLT/FORS1 & SNIa  &  0.568 & gal\\
SNLS-04D2fp & 09:59:28.162 & +02:19:15.58 &   2641 & VLT/FORS1 & SNIa  &  0.415 & gal\\
SNLS-04D2fs & 10:00:22.110 & +01:45:55.70 &   2641 & VLT/FORS1 & SNIa  &  0.357 & gal\\
SNLS-04D2gb & 10:02:22.676 & +01:53:39.34 &   3117 & Keck/LRIS & SNIa &  0.43 & SN\\
SNLS-04D2gc & 10:01:39.281 & +01:52:59.36 &   3118 & Keck/LRIS & SNIa &  0.521 & gal\\
SNLS-04D2gp & 09:59:20.400 & +02:30:31.88 &   3116 & VLT/FORS1 & SNIa  &  0.71 & SN\\
SNLS-04D2iu & 10:01:13.221 & +02:24:53.91 &   3139 & VLT/FORS1 & SNIa*  &  0.69 & SN\\
SNLS-04D2ja & 09:58:48.519 & +01:46:18.64 &   3139 & VLT/FORS1 & SNIa*  &  0.74 & SN\\
SNLS-04D3bf & 14:17:45.096 & +52:28:04.31 &   3054 & Gem/GMOS & SNIa &  0.156 & gal\\
SNLS-04D3co & 14:17:50.030 & +52:57:48.95 &   3117 & Keck/LRIS & SNIa &  0.62 & SN\\
SNLS-04D3cp & 14:20:23.954 & +52:49:15.45 &   3119 & Keck/LRIS & SNIa &  0.83 & SN\\
SNLS-04D3cy & 14:18:12.452 & +52:39:30.40 &   3115 & Keck/DMOS & SNIa &  0.643 & gal\\
SNLS-04D3dd & 14:17:48.411 & +52:28:14.57 &   3122 & Gem/GMOS & SNIa &  1.01 & SN\\
SNLS-04D3df & 14:18:10.042 & +52:16:39.85 &   3117 & Keck/LRIS & SNIa &  0.47 & SN\\
SNLS-04D3do & 14:17:46.113 & +52:16:03.36 &   3117 & Keck/LRIS & SNIa &  0.61 & SN\\
SNLS-04D3ez & 14:19:07.894 & +53:04:19.17 &   3118 & Keck/LRIS & SNIa &  0.263 & gal\\
SNLS-04D3fk & 14:18:26.198 & +52:31:42.74 &   3118 & Keck/LRIS & SNIa &  0.358 & gal\\
SNLS-04D3fq & 14:16:57.902 & +52:22:46.46 &   3123 & Gem/GMOS & SNIa &  0.73 & SN\\
SNLS-04D3gt & 14:22:32.611 & +52:38:49.30 &   3149 & Keck/LRIS & SNIa* &  0.451 & gal\\
SNLS-04D3gx & 14:20:13.666 & +52:16:58.33 &   3147 & Gem/GMOS & SNIa* &  0.91 & SN\\
SNLS-04D3hn & 14:22:06.908 & +52:13:43.00 &   3148 & Gem/GMOS & SNIa &  0.552 & gal\\
SNLS-04D3is & 14:16:51.968 & +52:48:45.70 &   3149 & Keck/LRIS & SNIa* &  0.71 & SN\\
SNLS-04D3ki & 14:19:34.598 & +52:17:32.61 &   3149 & Keck/LRIS & SNIa* &  0.930 & gal\\
SNLS-04D3kr & 14:16:35.943 & +52:28:44.02 &   3173 & Gem/GMOS & SNIa &  0.337 & gal\\
SNLS-04D3ks & 14:22:33.479 & +52:11:07.44 &   3149 & Keck/LRIS & SNIa* &  0.752 & gal\\
SNLS-04D3lp & 14:19:50.911 & +52:30:11.88 &   3153 & Gem/GMOS & SNIa* &  0.983 & gal\\
SNLS-04D3lu & 14:21:08.009 & +52:58:29.74 &   3180 & Gem/GMOS & SNIa &  0.822 & gal\\
SNLS-04D3mk & 14:19:25.768 & +53:09:49.48 &   3176 & Gem/GMOS & SNIa &  0.813 & gal\\
SNLS-04D3ml & 14:16:39.095 & +53:05:35.89 &   3177 & Gem/GMOS & SNIa &  0.95 & SN\\
SNLS-04D3nc & 14:16:18.224 & +52:16:26.09 &   3200 & Gem/GMOS & SNIa* &  0.817 & gal\\
SNLS-04D3nh & 14:22:26.729 & +52:20:00.92 &   3180 & Gem/GMOS & SNIa &  0.340 & gal\\
SNLS-04D3nq & 14:20:19.193 & +53:09:15.90 &   3201 & Gem/GMOS & SNIa &  0.22 & SN\\
SNLS-04D3nr & 14:22:38.526 & +52:38:55.89 &   3202 & Gem/GMOS & SNIa* &  0.96 & SN\\
SNLS-04D3ny & 14:18:56.332 & +52:11:15.06 &   3197 & Gem/GMOS & SNIa &  0.81 & SN\\
SNLS-04D3oe & 14:19:39.381 & +52:33:14.21 &   3198 & Gem/GMOS & SNIa &  0.756 & gal\\
SNLS-04D4an & 22:15:57.119 & -17:41:43.93 &   3200 & VLT/FORS1 & SNIa  &  0.613 & gal\\
SNLS-04D4bk & 22:15:07.681 & -18:03:36.79 &   3200 & VLT/FORS1 & SNIa*  &  0.84 & SN\\
SNLS-04D4bq & 22:14:49.391 & -17:49:39.37 &   3203 & VLT/FORS1 & SNIa  &  0.55 & SN\\
SNLS-04D4dm & 22:15:25.470 & -17:14:42.71 &   3206 & Gem/GMOS & SNIa &  0.811 & gal\\
SNLS-04D4dw & 22:16:44.667 & -17:50:02.38 &   3206 & VLT/FORS1 & SNIa  &  0.96 & SN\\

\enddata
\tablenotetext{a}{Date of spectrocopic observations (JD 2450000+).}
\tablenotetext{b}{Telescope and instrument with which the spectrum was acquired.}
\tablenotetext{c}{See Sect. \ref{section:spectro_follow_up} for definitions.}
\tablenotetext{d}{SN spectrum (SN) or host galaxy spectrum (gal). $\delta{z}\sim0.01$ when from SN spectrum, 
$\sim0.001$ when from host galaxy spectrum}  
\end{deluxetable}

%% file: nearbydata.tex

\begin{deluxetable}{lcccccccc}
\tablewidth{0pt}
\tablecaption{Nearby Type Ia supernovae\label{table:nearbydata}}
\tablehead{
\colhead{Name} & 
\colhead{$z$ \tablenotemark{a}} & 
\colhead{Bands} &
\colhead{$m^*_B$} & 
\colhead{$s$} &
\colhead{$c$} &
\colhead{$\mu_B$ \tablenotemark{b}} &
\colhead{Phot. Ref.\tablenotemark{c}}  
}
\startdata
1990af & 0.050 & BV & 17.723 $\pm$ 0.006 & 0.737 $\pm$ 0.001 & -0.001 $\pm$  0.009 & 36.632 $\pm$  0.045 & (H96) \\
1990O & 0.031 & BV & 16.196 $\pm$ 0.023\tablenotemark{d} & 1.035 $\pm$ 0.033\tablenotemark{d} & 0.017 $\pm$  0.023\tablenotemark{d} & 35.532 $\pm$  0.091 & (H96) \\
1992ae & 0.075 & BV & 18.392 $\pm$ 0.037\tablenotemark{d} & 0.939 $\pm$ 0.021\tablenotemark{d} & -0.023 $\pm$  0.025\tablenotemark{d} & 37.642 $\pm$  0.049 & (H96) \\
1992ag & 0.026 & BV & 16.241 $\pm$ 0.021\tablenotemark{d} & 1.030 $\pm$ 0.027\tablenotemark{d} & 0.155 $\pm$  0.018\tablenotemark{d} & 35.353 $\pm$  0.094 & (H96) \\
1992aq & 0.101 & BV & 19.299 $\pm$ 0.028\tablenotemark{d} & 0.839 $\pm$ 0.032\tablenotemark{d} & -0.048 $\pm$  0.020\tablenotemark{d} & 38.437 $\pm$  0.055 & (H96) \\
1992bc & 0.020 & BV & 15.086 $\pm$ 0.007 & 1.033 $\pm$ 0.007 & -0.031 $\pm$  0.008 & 34.494 $\pm$  0.111 & (H96) \\
1992bh & 0.045 & BV & 17.592 $\pm$ 0.016 & 0.985 $\pm$ 0.016 & 0.095 $\pm$  0.014 & 36.728 $\pm$  0.057 & (H96) \\
1992bl & 0.043 & BV & 17.275 $\pm$ 0.033\tablenotemark{d} & 0.784 $\pm$ 0.016\tablenotemark{d} & -0.014 $\pm$  0.020\tablenotemark{d} & 36.276 $\pm$  0.059 & (H96) \\
1992bo & 0.018 & BV & 15.753 $\pm$ 0.012 & 0.739 $\pm$ 0.006 & 0.055 $\pm$  0.011 & 34.576 $\pm$  0.121 & (H96) \\
1992bp & 0.079 & BV & 18.281 $\pm$ 0.011 & 0.873 $\pm$ 0.014 & -0.043 $\pm$  0.012 & 37.465 $\pm$  0.041 & (H96) \\
1992br & 0.088 & BV & 19.398 $\pm$ 0.073\tablenotemark{d} & 0.650 $\pm$ 0.029\tablenotemark{d} & 0.032 $\pm$  0.037\tablenotemark{d} & 38.121 $\pm$  0.046 & (H96) \\
1992bs & 0.063 & BV & 18.177 $\pm$ 0.041\tablenotemark{d} & 1.001 $\pm$ 0.018\tablenotemark{d} & -0.034 $\pm$  0.019\tablenotemark{d} & 37.540 $\pm$  0.046 & (H96) \\
1992P & 0.026 & BV & 16.037 $\pm$ 0.018\tablenotemark{d} & 1.139 $\pm$ 0.084\tablenotemark{d} & -0.005 $\pm$  0.018\tablenotemark{d} & 35.565 $\pm$  0.141 & (H96) \\
1993ag & 0.050 & BV & 17.799 $\pm$ 0.014 & 0.915 $\pm$ 0.018 & 0.096 $\pm$  0.017 & 36.827 $\pm$  0.060 & (H96) \\
1993B & 0.071 & BV & 18.377 $\pm$ 0.054\tablenotemark{d} & 0.988 $\pm$ 0.022\tablenotemark{d} & 0.041 $\pm$  0.026\tablenotemark{d} & 37.604 $\pm$  0.048 & (H96) \\
1993H & 0.025 & BV & 16.735 $\pm$ 0.017 & 0.699 $\pm$ 0.012 & 0.250 $\pm$  0.015 & 35.192 $\pm$  0.092 & (H96,A04) \\
1993O & 0.053 & BV & 17.614 $\pm$ 0.011 & 0.901 $\pm$ 0.010 & -0.014 $\pm$  0.011 & 36.794 $\pm$  0.047 & (H96) \\
1994M & 0.024 & BV & 16.205 $\pm$ 0.041\tablenotemark{d} & 0.854 $\pm$ 0.019\tablenotemark{d} & 0.040 $\pm$  0.022\tablenotemark{d} & 35.228 $\pm$  0.094 & (R99) \\
1994S & 0.016 & BV & 14.760 $\pm$ 0.017 & 1.018 $\pm$ 0.026 & 0.016 $\pm$  0.017 & 34.071 $\pm$  0.146 & (R99) \\
1995ac & 0.049 & BV & 17.026 $\pm$ 0.009 & 1.042 $\pm$ 0.013 & 0.010 $\pm$  0.010 & 36.383 $\pm$  0.051 & (R99,A04) \\
1995bd & 0.016 & BV & 15.246 $\pm$ 0.009 & 0.992 $\pm$ 0.009 & 0.293 $\pm$  0.008 & 34.083 $\pm$  0.138 & (R99,A04) \\
1996ab & 0.125 & BV & 19.525 $\pm$ 0.027\tablenotemark{d} & 0.957 $\pm$ 0.033\tablenotemark{d} & -0.074 $\pm$  0.015\tablenotemark{d} & 38.885 $\pm$  0.049 & (R99) \\
1996bl & 0.035 & BV & 16.611 $\pm$ 0.010 & 0.983 $\pm$ 0.015 & 0.037 $\pm$  0.011 & 35.837 $\pm$  0.069 & (R99) \\
1996bo & 0.016 & UBV & 15.816 $\pm$ 0.006 & 0.881 $\pm$ 0.003 & 0.343 $\pm$  0.007 & 34.405 $\pm$  0.133 & (R99,A04) \\
1996bv & 0.017 & BV & 15.380 $\pm$ 0.019\tablenotemark{d} & 0.989 $\pm$ 0.024\tablenotemark{d} & 0.225 $\pm$  0.009\tablenotemark{d} & 34.319 $\pm$  0.133 & (R99) \\
1996C & 0.030 & BV & 16.636 $\pm$ 0.029\tablenotemark{d} & 1.045 $\pm$ 0.111\tablenotemark{d} & 0.122 $\pm$  0.010\tablenotemark{d} & 35.822 $\pm$  0.210 & (R99) \\
1997dg & 0.030 & UBV & 16.821 $\pm$ 0.014\tablenotemark{d} & 0.917 $\pm$ 0.024\tablenotemark{d} & 0.005 $\pm$  0.010\tablenotemark{d} & 35.994 $\pm$  0.080 & (J02) \\
1997Y & 0.017 & UBV & 15.284 $\pm$ 0.020\tablenotemark{d} & 0.916 $\pm$ 0.024\tablenotemark{d} & 0.008 $\pm$  0.014\tablenotemark{d} & 34.452 $\pm$  0.136 & (J02) \\
1998ab & 0.028 & UBV & 16.048 $\pm$ 0.010 & 0.938 $\pm$ 0.008 & 0.071 $\pm$  0.007 & 35.150 $\pm$  0.079 & (J02) \\
1998dx & 0.054 & UBV & 17.660 $\pm$ 0.055\tablenotemark{d} & 0.733 $\pm$ 0.039\tablenotemark{d} & -0.028 $\pm$  0.019\tablenotemark{d} & 36.606 $\pm$  0.054 & (J02) \\
1998eg & 0.024 & UBV & 16.089 $\pm$ 0.009 & 0.940 $\pm$ 0.029 & 0.036 $\pm$  0.012 & 35.250 $\pm$  0.102 & (J02) \\
1998V & 0.017 & UBV & 15.094 $\pm$ 0.011\tablenotemark{d} & 0.909 $\pm$ 0.016\tablenotemark{d} & 0.030 $\pm$  0.006\tablenotemark{d} & 34.216 $\pm$  0.128 & (J02) \\
1999aw & 0.039 & BV & 16.732 $\pm$ 0.005 & 1.205 $\pm$ 0.008 & 0.044 $\pm$  0.006 & 36.284 $\pm$  0.057 & (S02) \\
1999cc & 0.032 & UBV & 16.791 $\pm$ 0.009 & 0.840 $\pm$ 0.013 & 0.043 $\pm$  0.010 & 35.789 $\pm$  0.074 & (J02) \\
1999ek & 0.018 & UBV & 15.584 $\pm$ 0.004 & 0.892 $\pm$ 0.007 & 0.153 $\pm$  0.005 & 34.489 $\pm$  0.124 & (J02,K04b) \\
1999gp & 0.026 & UBV & 16.005 $\pm$ 0.004 & 1.104 $\pm$ 0.007 & 0.083 $\pm$  0.004 & 35.342 $\pm$  0.084 & (J02,K01) \\
2000ca & 0.025 & UBV & 15.510 $\pm$ 0.007 & 1.006 $\pm$ 0.013 & -0.066 $\pm$  0.006 & 34.931 $\pm$  0.091 & (K04a) \\
2000cf & 0.036 & UBV & 17.091 $\pm$ 0.027\tablenotemark{d} & 0.868 $\pm$ 0.024\tablenotemark{d} & 0.054 $\pm$  0.013\tablenotemark{d} & 36.113 $\pm$  0.066 & (J02) \\
2000cn & 0.023 & UBV & 16.544 $\pm$ 0.007 & 0.732 $\pm$ 0.006 & 0.190 $\pm$  0.006 & 35.146 $\pm$  0.094 & (J02) \\
2000dk & 0.016 & UBV & 15.323 $\pm$ 0.005 & 0.724 $\pm$ 0.006 & 0.052 $\pm$  0.005 & 34.129 $\pm$  0.133 & (J02) \\
2000fa & 0.022 & UBV & 15.832 $\pm$ 0.014 & 0.953 $\pm$ 0.010 & 0.081 $\pm$  0.009 & 34.941 $\pm$  0.101 & (J02) \\
2001ba & 0.031 & BV & 16.182 $\pm$ 0.006 & 1.000 $\pm$ 0.011 & -0.043 $\pm$  0.008 & 35.558 $\pm$  0.075 & (K04a) \\
2001cn & 0.015 & UBV & 15.271 $\pm$ 0.013\tablenotemark{d} & 0.911 $\pm$ 0.012\tablenotemark{d} & 0.208 $\pm$  0.007\tablenotemark{d} & 34.118 $\pm$  0.142 & (K04b) \\
2001cz & 0.017 & UBV & 15.035 $\pm$ 0.006 & 1.004 $\pm$ 0.010 & 0.120 $\pm$  0.007 & 34.162 $\pm$  0.127 & (K04b) \\

\enddata
\tablenotetext{a}{CMB-centric redshift.}
\tablenotetext{b}{Computed with $H_0 = 70$ km\,s$^{-1}$Mpc$^{-1}$. Uncertainty only accounts for photometric uncertainties.}
\tablenotetext{c}{Photometry References : 
H96: \citet{Hamuy96b}, 
R99: \citet{Riess99a}, 
K01: \citet{Krisciunas01}, 
J02: \citet{Jha02}, 
A04: \citet{Altavilla04},  
K04a: \citet{Krisciunas04a}, 
K04b: \citet{Krisciunas04b},
S02:\citet{Strolger02}.
}
\tablenotetext{d}{First photometric measurement after $B$-band maximum, see discussion in section~\ref{section:snIa-sample}}
\end{deluxetable}

%% file: snlsdata.tex

\begin{deluxetable}{lcccccc}
\tablewidth{0pt}
\tablecaption{SNLS Type Ia supernovae\label{table:snlsdata}}
\tablehead{
\colhead{Name} & 
\colhead{$z$ \tablenotemark{a}} & 
\colhead{Bands} &
\colhead{$m^*_B$} & 
\colhead{stretch\tablenotemark{b}} &
\colhead{color\tablenotemark{b}} &
\colhead{$\mu_B$\tablenotemark{c}} 
}
\startdata
SNLS-03D1au & 0.504  & riz & 22.978 $\pm$ 0.010 & 1.124 $\pm$ 0.019 & 0.030 $\pm$  0.018 & 42.429 $\pm$  0.039  \\
SNLS-03D1aw & 0.582  & riz & 23.599 $\pm$ 0.020 & 1.002 $\pm$ 0.024 & 0.018 $\pm$  0.030 & 42.881 $\pm$  0.054  \\
SNLS-03D1ax & 0.496  & riz & 22.957 $\pm$ 0.011 & 0.899 $\pm$ 0.010 & -0.044 $\pm$  0.021 & 42.180 $\pm$  0.038  \\
SNLS-03D1bp & 0.346  & riz & 22.465 $\pm$ 0.014 & 0.880 $\pm$ 0.007 & 0.143 $\pm$  0.017 & 41.367 $\pm$  0.021  \\
SNLS-03D1cm & 0.870  & griz & 24.469 $\pm$ 0.066 & 1.173 $\pm$ 0.061 & -0.035 $\pm$  0.143 & 44.095 $\pm$  0.301  \\
SNLS-03D1co & 0.679  & griz & 24.094 $\pm$ 0.033 & 0.975 $\pm$ 0.032 & -0.021 $\pm$  0.047 & 43.398 $\pm$  0.088  \\
SNLS-03D1ew & 0.868  & griz & 24.359 $\pm$ 0.078 & 1.028 $\pm$ 0.040 & -0.102 $\pm$  0.169 & 43.871 $\pm$  0.344  \\
SNLS-03D1fc & 0.331  & griz & 21.800 $\pm$ 0.005 & 0.937 $\pm$ 0.005 & 0.042 $\pm$  0.004 & 40.946 $\pm$  0.013  \\
SNLS-03D1fl & 0.688  & griz & 23.629 $\pm$ 0.015 & 0.999 $\pm$ 0.024 & -0.070 $\pm$  0.021 & 43.046 $\pm$  0.049  \\
SNLS-03D1fq & 0.800  & griz & 24.519 $\pm$ 0.030 & 0.806 $\pm$ 0.052 & 0.027 $\pm$  0.030 & 43.490 $\pm$  0.090  \\
SNLS-03D1gt & 0.548  & griz & 24.119 $\pm$ 0.048 & 0.856 $\pm$ 0.042 & 0.244 $\pm$  0.050 & 42.825 $\pm$  0.080  \\
SNLS-03D3af & 0.532  & gri & 23.470 $\pm$ 0.027 & 0.907 $\pm$ 0.023 & 0.029 $\pm$  0.031 & 42.592 $\pm$  0.083  \\
SNLS-03D3aw & 0.449  & griz & 22.552 $\pm$ 0.016 & 0.955 $\pm$ 0.013 & -0.048 $\pm$  0.019 & 41.866 $\pm$  0.044  \\
SNLS-03D3ay & 0.371  & griz & 22.201 $\pm$ 0.016 & 0.968 $\pm$ 0.010 & -0.018 $\pm$  0.014 & 41.488 $\pm$  0.030  \\
SNLS-03D3ba & 0.291  & griz & 22.049 $\pm$ 0.034 & 1.036 $\pm$ 0.021 & 0.263 $\pm$  0.015 & 40.999 $\pm$  0.033  \\
SNLS-03D3bh & 0.249  & griz & 21.132 $\pm$ 0.018 & 0.993 $\pm$ 0.008 & -0.090 $\pm$  0.013 & 40.571 $\pm$  0.020  \\
SNLS-03D3cc & 0.463  & gri & 22.558 $\pm$ 0.111 & 1.074 $\pm$ 0.031 & -0.070 $\pm$  0.050 & 42.089 $\pm$  0.034  \\
SNLS-03D3cd & 0.461  & gri & 22.562 $\pm$ 0.017 & 1.131 $\pm$ 0.034 & 0.025 $\pm$  0.011 & 42.031 $\pm$  0.058  \\
SNLS-03D4ag & 0.285  & griz & 21.237 $\pm$ 0.005 & 1.059 $\pm$ 0.005 & -0.061 $\pm$  0.004 & 40.731 $\pm$  0.015  \\
SNLS-03D4at & 0.633  & griz & 23.746 $\pm$ 0.020 & 0.989 $\pm$ 0.029 & -0.060 $\pm$  0.030 & 43.133 $\pm$  0.064  \\
SNLS-03D4au\tablenotemark{d} & 0.468  & griz & 23.856 $\pm$ 0.020 & 1.000 $\pm$ 0.030 & 0.291 $\pm$  0.034 & 42.708 $\pm$  0.069  \\
SNLS-03D4bc\tablenotemark{d} & 0.572  & griz & 24.596 $\pm$ 0.061 & 0.774 $\pm$ 0.048 & 0.025 $\pm$  0.078 & 43.521 $\pm$  0.135  \\
SNLS-03D4cn & 0.818  & griz & 24.652 $\pm$ 0.051 & 0.743 $\pm$ 0.059 & 0.023 $\pm$  0.158 & 43.532 $\pm$  0.304  \\
SNLS-03D4cx & 0.949  & griz & 24.504 $\pm$ 0.083 & 0.882 $\pm$ 0.019 & 0.080 $\pm$  0.124 & 43.507 $\pm$  0.272  \\
SNLS-03D4cy & 0.927  & griz & 24.718 $\pm$ 0.109 & 1.031 $\pm$ 0.052 & -0.305 $\pm$  0.174 & 44.553 $\pm$  0.380  \\
SNLS-03D4cz & 0.695  & griz & 24.019 $\pm$ 0.036 & 0.729 $\pm$ 0.024 & -0.069 $\pm$  0.043 & 43.023 $\pm$  0.086  \\
SNLS-03D4dh & 0.627  & griz & 23.389 $\pm$ 0.011 & 1.061 $\pm$ 0.013 & 0.028 $\pm$  0.016 & 42.746 $\pm$  0.035  \\
SNLS-03D4di & 0.905  & griz & 24.288 $\pm$ 0.068 & 1.103 $\pm$ 0.041 & 0.029 $\pm$  0.120 & 43.708 $\pm$  0.258  \\
SNLS-03D4dy & 0.604  & griz & 23.313 $\pm$ 0.010 & 1.056 $\pm$ 0.001 & 0.122 $\pm$  0.015 & 42.515 $\pm$  0.029  \\
SNLS-03D4fd & 0.791  & griz & 24.212 $\pm$ 0.025 & 0.919 $\pm$ 0.033 & 0.028 $\pm$  0.044 & 43.353 $\pm$  0.076  \\
SNLS-03D4gf & 0.581  & griz & 23.351 $\pm$ 0.013 & 1.009 $\pm$ 0.026 & -0.056 $\pm$  0.024 & 42.761 $\pm$  0.047  \\
SNLS-03D4gg & 0.592  & griz & 23.403 $\pm$ 0.024 & 0.966 $\pm$ 0.049 & 0.062 $\pm$  0.035 & 42.562 $\pm$  0.090  \\
SNLS-03D4gl & 0.571  & griz & 23.269 $\pm$ 0.026 & 0.957 $\pm$ 0.033 & 0.030 $\pm$  0.028 & 42.465 $\pm$  0.070  \\
SNLS-04D1ag & 0.557  & griz & 23.003 $\pm$ 0.011 & 0.944 $\pm$ 0.013 & -0.182 $\pm$  0.017 & 42.511 $\pm$  0.029  \\
SNLS-04D1aj & 0.721  & griz & 23.901 $\pm$ 0.030 & 1.074 $\pm$ 0.067 & 0.072 $\pm$  0.038 & 43.209 $\pm$  0.106  \\
SNLS-04D1ak & 0.526  & griz & 23.631 $\pm$ 0.028 & 0.824 $\pm$ 0.021 & 0.018 $\pm$  0.033 & 42.644 $\pm$  0.055  \\
SNLS-04D2cf & 0.369  & griz & 22.340 $\pm$ 0.007 & 0.895 $\pm$ 0.003 & 0.002 $\pm$  0.010 & 41.485 $\pm$  0.016  \\
SNLS-04D2fp & 0.415  & griz & 22.528 $\pm$ 0.010 & 0.964 $\pm$ 0.010 & 0.006 $\pm$  0.015 & 41.772 $\pm$  0.027  \\
SNLS-04D2fs & 0.357  & griz & 22.422 $\pm$ 0.008 & 0.942 $\pm$ 0.009 & 0.128 $\pm$  0.008 & 41.441 $\pm$  0.018  \\
SNLS-04D2gb & 0.430  & griz & 22.796 $\pm$ 0.018 & 0.777 $\pm$ 0.013 & -0.008 $\pm$  0.025 & 41.776 $\pm$  0.038  \\
SNLS-04D2gc & 0.521  & griz & 23.321 $\pm$ 0.014 & 1.065 $\pm$ 0.024 & 0.185 $\pm$  0.022 & 42.439 $\pm$  0.054  \\
SNLS-04D2gp & 0.707  & griz & 24.151 $\pm$ 0.047 & 0.801 $\pm$ 0.002 & -0.052 $\pm$  0.060 & 43.237 $\pm$  0.129  \\
SNLS-04D2iu & 0.691  & griz & 24.258 $\pm$ 0.048 & 0.800 $\pm$ 0.035 & 0.074 $\pm$  0.056 & 43.144 $\pm$  0.136  \\
SNLS-04D2ja & 0.741  & griz & 24.098 $\pm$ 0.045 & 0.945 $\pm$ 0.036 & -0.067 $\pm$  0.043 & 43.427 $\pm$  0.117  \\
SNLS-04D3co & 0.620  & griz & 23.781 $\pm$ 0.022 & 0.895 $\pm$ 0.017 & -0.064 $\pm$  0.030 & 43.030 $\pm$  0.060  \\
SNLS-04D3cp & 0.830  & griz & 24.235 $\pm$ 0.063 & 1.110 $\pm$ 0.035 & -0.448 $\pm$  0.180 & 44.414 $\pm$  0.347  \\
SNLS-04D3cy & 0.643  & griz & 23.798 $\pm$ 0.021 & 0.963 $\pm$ 0.016 & 0.017 $\pm$  0.029 & 43.023 $\pm$  0.059  \\
SNLS-04D3dd & 1.010  & griz & 25.120 $\pm$ 0.192 & 1.088 $\pm$ 0.074 & -0.071 $\pm$  0.205 & 44.673 $\pm$  0.533  \\
SNLS-04D3df & 0.470  & griz & 23.465 $\pm$ 0.010 & 0.730 $\pm$ 0.010 & 0.060 $\pm$  0.017 & 42.268 $\pm$  0.032  \\
SNLS-04D3do & 0.610  & griz & 23.574 $\pm$ 0.014 & 0.862 $\pm$ 0.013 & -0.079 $\pm$  0.019 & 42.796 $\pm$  0.039  \\
SNLS-04D3ez & 0.263  & griz & 21.678 $\pm$ 0.004 & 0.895 $\pm$ 0.006 & 0.091 $\pm$  0.003 & 40.682 $\pm$  0.013  \\
SNLS-04D3fk & 0.358  & griz & 22.532 $\pm$ 0.005 & 0.913 $\pm$ 0.005 & 0.149 $\pm$  0.006 & 41.474 $\pm$  0.013  \\
SNLS-04D3fq & 0.730  & griz & 24.128 $\pm$ 0.026 & 0.900 $\pm$ 0.014 & -0.002 $\pm$  0.037 & 43.287 $\pm$  0.075  \\
SNLS-04D3gt & 0.451  & griz & 23.235 $\pm$ 0.010 & 0.953 $\pm$ 0.010 & 0.276 $\pm$  0.016 & 42.038 $\pm$  0.030  \\
SNLS-04D3gx & 0.910  & griz & 24.708 $\pm$ 0.094 & 0.952 $\pm$ 0.047 & -0.202 $\pm$  0.163 & 44.259 $\pm$  0.346  \\
SNLS-04D3hn & 0.552  & griz & 23.475 $\pm$ 0.011 & 0.898 $\pm$ 0.011 & 0.106 $\pm$  0.017 & 42.461 $\pm$  0.035  \\
SNLS-04D3is & 0.710  & griz & 24.256 $\pm$ 0.027 & 0.972 $\pm$ 0.002 & 0.220 $\pm$  0.038 & 43.176 $\pm$  0.077  \\
SNLS-04D3ki & 0.930  & griz & 24.871 $\pm$ 0.126 & 0.901 $\pm$ 0.039 & -0.256 $\pm$  0.194 & 44.430 $\pm$  0.430  \\
SNLS-04D3kr & 0.337  & griz & 21.967 $\pm$ 0.003 & 1.064 $\pm$ 0.004 & 0.072 $\pm$  0.003 & 41.259 $\pm$  0.010  \\
SNLS-04D3ks & 0.752  & griz & 23.882 $\pm$ 0.035 & 1.013 $\pm$ 0.037 & 0.026 $\pm$  0.043 & 43.170 $\pm$  0.090  \\
SNLS-04D3lp & 0.983  & griz & 24.925 $\pm$ 0.168 & 0.831 $\pm$ 0.049 & 0.022 $\pm$  0.211 & 43.941 $\pm$  0.496  \\
SNLS-04D3lu & 0.822  & griz & 24.342 $\pm$ 0.040 & 0.950 $\pm$ 0.028 & 0.019 $\pm$  0.116 & 43.544 $\pm$  0.218  \\
SNLS-04D3ml & 0.950  & griz & 24.552 $\pm$ 0.082 & 1.182 $\pm$ 0.015 & 0.117 $\pm$  0.122 & 43.954 $\pm$  0.268  \\
SNLS-04D3nc & 0.817  & griz & 24.271 $\pm$ 0.048 & 1.111 $\pm$ 0.064 & 0.062 $\pm$  0.140 & 43.652 $\pm$  0.254  \\
SNLS-04D3nh & 0.340  & griz & 22.137 $\pm$ 0.004 & 1.011 $\pm$ 0.006 & 0.089 $\pm$  0.004 & 41.323 $\pm$  0.012  \\
SNLS-04D3nr & 0.960  & griz & 24.542 $\pm$ 0.075 & 0.922 $\pm$ 0.045 & 0.070 $\pm$  0.110 & 43.622 $\pm$  0.234  \\
SNLS-04D3ny & 0.810  & griz & 24.272 $\pm$ 0.050 & 1.005 $\pm$ 0.084 & -0.065 $\pm$  0.152 & 43.691 $\pm$  0.301  \\
SNLS-04D3oe & 0.756  & griz & 24.069 $\pm$ 0.026 & 0.783 $\pm$ 0.028 & -0.259 $\pm$  0.033 & 43.453 $\pm$  0.058  \\
SNLS-04D4an & 0.613  & griz & 24.022 $\pm$ 0.023 & 0.823 $\pm$ 0.025 & 0.064 $\pm$  0.025 & 42.961 $\pm$  0.061  \\
SNLS-04D4bk & 0.840  & griz & 24.314 $\pm$ 0.037 & 1.050 $\pm$ 0.051 & 0.142 $\pm$  0.098 & 43.475 $\pm$  0.185  \\
SNLS-04D4bq & 0.550  & griz & 23.362 $\pm$ 0.020 & 0.995 $\pm$ 0.029 & 0.112 $\pm$  0.027 & 42.487 $\pm$  0.056  \\
SNLS-04D4dm & 0.811  & griz & 24.390 $\pm$ 0.044 & 1.000 $\pm$ 0.057 & -0.161 $\pm$  0.150 & 43.950 $\pm$  0.264  \\
SNLS-04D4dw & 0.961  & griz & 24.566 $\pm$ 0.093 & 0.962 $\pm$ 0.058 & -0.117 $\pm$  0.138 & 44.000 $\pm$  0.290  \\

\enddata
\tablenotetext{a}{Heliocentric redshift.}
\tablenotetext{b}{See section \ref{section:lightcurve_fit_model} for description.}
\tablenotetext{c}{Computed with $H_0 = 70$ km\,s$^{-1}$Mpc$^{-1}$. Uncertainty only accounts for photometric uncertainties.}
\tablenotetext{d}{Not included in the final cosmological fits.}
\end{deluxetable}